\documentclass[journal=jacsat,manuscript=article]{achemso}

\usepackage{color}
\usepackage{soul}
\usepackage{graphicx}
\usepackage{amsfonts}
\usepackage{amsmath,amssymb}

\usepackage{xy}
\usepackage{xr}
\usepackage{setspace}

\usepackage[normalem]{ulem}

\makeatletter
\newcommand*{\addFileDependency}[1]{
  \typeout{(#1)}
  \@addtofilelist{#1}
  \IfFileExists{#1}{}{\typeout{No file #1.}}
}
\makeatother

\title{Multiscale Growth Kinetics of Model Biomolecular Condensates Under Passive and Active Conditions}

\author{Tamizhmalar Sundararajan}
\affiliation{Universit\'e Paris-Saclay, CNRS, Laboratoire de Physique des Solides, 91405 Orsay, France}
\author{Matteo Boccalini}
\affiliation{Centre de Biologie Structurale, Université de Montpellier, CNRS, INSERM, 34090 Montpellier, France}
\author{Roméo Suss}
\affiliation{Universit\'e Paris-Saclay, CNRS, Laboratoire de Physique des Solides, 91405 Orsay, France}
\alsoaffiliation{Universit\'e Paris-Saclay, ENS Paris-Saclay, CNRS, CentraleSupélec, LuMIn, 91190 Gif-sur-Yvette, France}
\author{Sandrine Mariot}
\affiliation{Universit\'e Paris-Saclay, CNRS, Laboratoire de Physique des Solides, 91405 Orsay, France}
\author{Emerson R. Da Silva}
\affiliation{Departamento de Biofísica, Universidade Federal de São Paulo, São Paulo 04062-000, Brazil}
\alsoaffiliation{Universit\'e Paris-Saclay, CNRS, Laboratoire de Physique des Solides, 91405 Orsay, France}
\author{Fernando C. Giacomelli}
\affiliation{Centro de Ciências Naturais e Humanas, Universidade Federal do ABC, Santo André, Brazil}
\alsoaffiliation{Universit\'e Paris-Saclay, CNRS, Laboratoire de Physique des Solides, 91405 Orsay, France}
\author{Austin Hubley}
\affiliation{ESRF - The European Synchrotron, 38043 Grenoble, France}
\author{Theyencheri Narayanan}
\affiliation{ESRF - The European Synchrotron, 38043 Grenoble, France}
\author{Alessandro Barducci}
\affiliation{Centre de Biologie Structurale, Université de Montpellier, CNRS, INSERM, 34090 Montpellier, France}
\author{Guillaume Tresset}
\affiliation{Universit\'e Paris-Saclay, CNRS, Laboratoire de Physique des Solides, 91405 Orsay, France}
\email{guillaume.tresset@universite-paris-saclay.fr}

\begin{document}

\begin{tocentry} 

\includegraphics{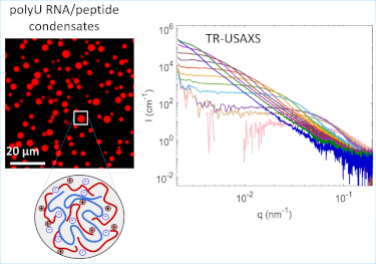}

\end{tocentry}

\begin{abstract}

Living cells exhibit a complex organization comprising numerous compartments, among which are RNA- and protein-rich membraneless, liquid-like organelles known as biomolecular condensates. Energy-consuming processes regulate their formation and dissolution, with (de-)phosphorylation by specific enzymes being among the most commonly involved reactions. By employing a model system consisting of a phosphorylatable peptide and homopolymeric RNA, we elucidate how enzymatic activity modulates the growth kinetics and alters the local structure of biomolecular condensates. Under passive condition, time-resolved ultra-small-angle X-ray scattering with synchrotron source reveals a nucleation-driven coalescence mechanism maintained over four decades in time, similar to the coarsening of simple binary fluid mixtures. Coarse-grained molecular dynamics simulations show that peptide-decorated RNA chains assembled shortly after mixing constitute the relevant subunits. In contrast, actively-formed condensates initially display a local mass fractal structure, which gradually matures upon enzymatic activity before condensates undergo coalescence. Both types of condensate eventually reach a steady state but fluorescence recovery after photobleaching indicates a peptide diffusivity twice higher in actively-formed condensates consistent with their loosely-packed local structure. We expect multiscale, integrative approaches implemented with model systems to link effectively the functional properties of membraneless organelles to their formation and dissolution kinetics as regulated by cellular active processes.
 
\end{abstract}

\section{Introduction}

Membraneless organelles are intracellular compartments devoid of lipid membrane, consisting of multiple proteins and possibly nucleic acids (DNA and/or RNA), and formed by liquid-liquid phase separation \cite{hyman_liquid-liquid_2014,shin_liquid_2017,mittag_conceptual_2022,hirose_guide_2023}. These biomolecular condensates perform essential biological functions by sequestering specific factors and promoting biochemical reactions following the needs of the cell \cite{lyon_framework_2021}. They also serve as protocells for the study of the emergence of functional cellular life or as catalytic microcompartments \cite{gao_membranized_2023,saini_biomolecular_2023}. Many condensates found in eucaryotic cells contain RNA and are involved in gene expression, such as nucleoli \cite{berry_rna_2015}, P granules \cite{brangwynne_germline_2009}, stress granules \cite{van_treeck_rna_2018} or Cajal bodies \cite{gall_centennial_2003} to name a few. Protein-RNA condensates may also play a role in pathological processes \cite{wang_liquidliquid_2021}, for example by facilitating the transcription of viral RNA during SARS-Cov-2 replication \cite{savastano_nucleocapsid_2020}. Membraneless organelles are relatively short-lived, in that they are formed to perform specific functions and are dissolved after the completion of these tasks. Such a flexibility is maintained through energy-consuming enzymatic processes. In particular, various post-translational modifications have been shown to regulate intracellular phase separation, including phosphorylation \cite{monahan_phosphorylation_2017}, methylation \cite{qamar_fus_2018}, acetylation and combinations thereof \cite{hofweber_friend_2019}.

Phase separation is a phenomenon by which a homogeneous material, spontaneously or not, separates into two or more distinct domains (phases) with different physical and/or chemical properties. Extensive experimental and theoretical efforts have been devoted to elucidating the phase-separation kinetics of binary systems \cite{bray_theory_2002}, including fluid mixtures relevant to biomolecular condensates, under both passive (spontaneous) and active (chemically driven) conditions \cite{cates_theories_2018,berry_physical_2018,zhou_fundamental_2024}. A major limitation in these investigations lies in the lack of information across length and time scales spanning several orders of magnitude. Indeed, biomolecular condensates are mostly micrometer-sized and therefore, fluorescence microscopy has become the standard technique to monitor their life cycle, especially within cells \cite{berry_rna_2015}. Quite recently, holographic microscopy has demonstrated its ability to monitor the growth of PopZ-cation condensates without the need for protein labeling \cite{von_hofe_multivalency_2025}. In all these cases, submicrometer-sized condensates formed along the growth pathway are overlooked, and events occurring on subsecond timescales often remain inaccessible. Conversely, coarse-grained molecular dynamics simulations can provide a wealth of information at the molecular level \cite{mathur_recent_2024}, and a few phase-separated protein-RNA condensates have been elucidated \cite{regy_sequence_2020,alshareedah_phase_2020}. Nonetheless, coarse-grained models face substantial challenges in quantitatively reproducing the thermodynamic properties of biomolecular condensates. In addition, although they allow the simulation of relatively large supramolecular systems, the accessible timescales remain limited to a few microseconds at most.

To bridge these multiple scales, time-resolved small-angle X-ray scattering (TR-SAXS) using synchrotron source has emerged as a state-of-the-art technique, enabling to track self-assembly processes with millisecond temporal resolution and spatial sensitivity ranging from a few to several hundred nanometers \cite{narayanan_recent_2017,chevreuil_nonequilibrium_2018,amann_kinetic_2019,dharan_hierarchical_2021,kra_energetics_2023}. Within the context of liquid-liquid phase separation, the assembly kinetics of a prion-like domain has revealed the existence of a double nucleation barrier to overcome before the condensate growth can proceed \cite{martin_multi-step_2021}. With the advent of time-resolved ultra-small-angle X-ray scattering (TR-USAXS) \cite{narayanan_performance_2022} available at a very few synchrotron facilities worldwide, the capabilities of X-ray scattering have extended to the micrometer scale. For example, the spinodal decomposition of bovine serum albumin induced by YCl\textsubscript{3} was probed during the first two minutes following the mixing of components \cite{vela_kinetics_2016}. Likewise, coacervates composed of oppositely-charged polyelectrolytes were found to change from mass fractal to surface fractal morphologies during their growth driven by coalescence \cite{takahashi_growth_2017}. Nevertheless, how enzymatic activity regulates the growth mechanism, the local structure, and the morphology of biomolecular condensates -- compared to conditions lacking active processes -- remains experimentally unaddressed.

To fill this knowledge gap, we employ a model system of biomolecular condensates whose formation is triggered by the dephosphorylation of serine residues, thereby mimicking the enzymatic regulation encountered in intracellular membraneless organelles \cite{aumiller_phosphorylation-mediated_2016}. This system consists of an arginine-rich peptide (RRASLRRASL) containing phosphorylatable serines, associated with poly(uridylic acid) (polyU) RNA. Condensates form spontaneously when the serines bear positive charges -- a condition hereafter referred to as passive -- or are enzymatically induced by lambda phosphatase (LPP), which removes phosphate groups from initially phosphorylated serines, defining the active condition. This flexible system enables the comparison between condensate formation mechanisms under passive and active conditions. We therefore combine TR-USAXS experiments performed at varying peptide and enzyme concentrations, with confocal microscopy and coarse-grained molecular dynamics simulations to capture these mechanisms across a wide range of length- and timescales.

\section{Results}

\subsection{Phase separation occurs in the cationic regime}

\begin{figure}[hbt!]
\centering
\includegraphics[width=0.5\linewidth]{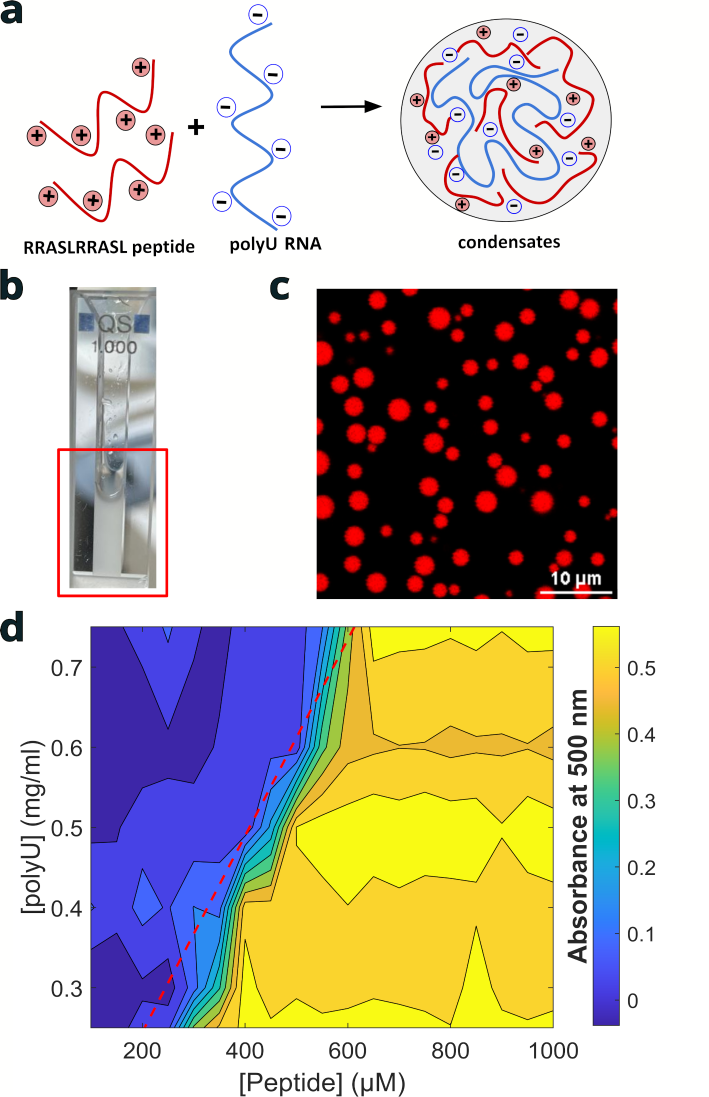}
\caption{(a) Schematic representation of RRASLRRASL peptide (red) and polyU RNA (blue) forming condensates upon Coulomb interaction. (b) A solution of condensates in a quartz cuvette exhibits a high level of turbidity. (c) Confocal microscopy image of condensates obtained with 1,000 µM of peptide including 0.5\% of TAMRA-peptide, and 0.5 g.L\textsuperscript{-1} of RNA. (d) Contour map of turbidity measured at 500 nm as a function of peptide and RNA concentrations. The red dashed line indicates electroneutrality.}
\label{fig:phase_diagram}
\end{figure}

Upon mixing RRASLRRASL peptide with anionic polyU RNA (Fig.~\ref{fig:phase_diagram}a) in 4 mM MgCl\textsubscript{2}, 2.25 mM ATP, 1.6 mM dithiothreitol, 50 mM HEPES pH 7.4 at 37 °C, biomolecular condensates in the form of liquid-like droplets are spontaneously made by coacervation \cite{aumiller_phosphorylation-mediated_2016}. A direct evidence for the presence of micrometer-sized droplets is given by the turbid aspect of the solution (Fig.~\ref{fig:phase_diagram}b). Confocal microscopy images performed with 0.5\% of fluorescent TAMRA-peptide confirm the spherical shape and the micrometer-scale size of the condensates (Fig.~\ref{fig:phase_diagram}c). At a given RNA concentration, turbidity measurements display a clear peptide threshold (saturation) concentration (Fig.~\ref{fig:phase_diagram}d), which is characteristic of phase separation. It is worth noting that the phase boundary is almost superimposed with the electroneutrality line. Phase separation occurs in the cationic regime, when the total positive charges supplied by the peptides outnumber the total negative charges carried by RNA.

The SAXS measurement of peptides reveals no sign of aggregation (Fig.~\ref{fig:SI-saxs_peptide}) given that the scattering curve at low $q$-values -- $q$ being the magnitude of the scattering wavevector -- is flat, which means that the Guinier plateau is reached. Furthermore, the measured radius of gyration ($R_\mathrm{g}=\text{0.9\ nm}$) is consistent with the small number of amino acids (ten). Likewise, the scattering curve of phosphorylated peptides reveals a radius of gyration of 0.84 nm (Fig.~\ref{fig:SI-saxs_peptide}). PolyU RNA has $250\pm80$ (s.d.) nucleotides according to mass photometry (Fig.~\ref{fig:SI-mass_photometry_polyU}) and its radius of gyration is estimated to be 11.6 nm (Fig.~\ref{fig:SI-saxs_polyU}). The Kratky representation of the RNA scattering curve displays an increase of the normalized intensity $(qR_\mathrm{g})^2(I/I_0)$ at high $qR_\mathrm{g}$ -- where $I$ stands for the scattering intensity and $I_0$ is the forward scattering intensity --, which indicates a stretched conformation as expected for a fully charged polyelectrolyte.

\subsection{Passively-formed condensates grow by nucleated Brownian motion-induced coalescence}

\begin{figure}[hbt!]
\centering
\includegraphics[width=0.75\linewidth]{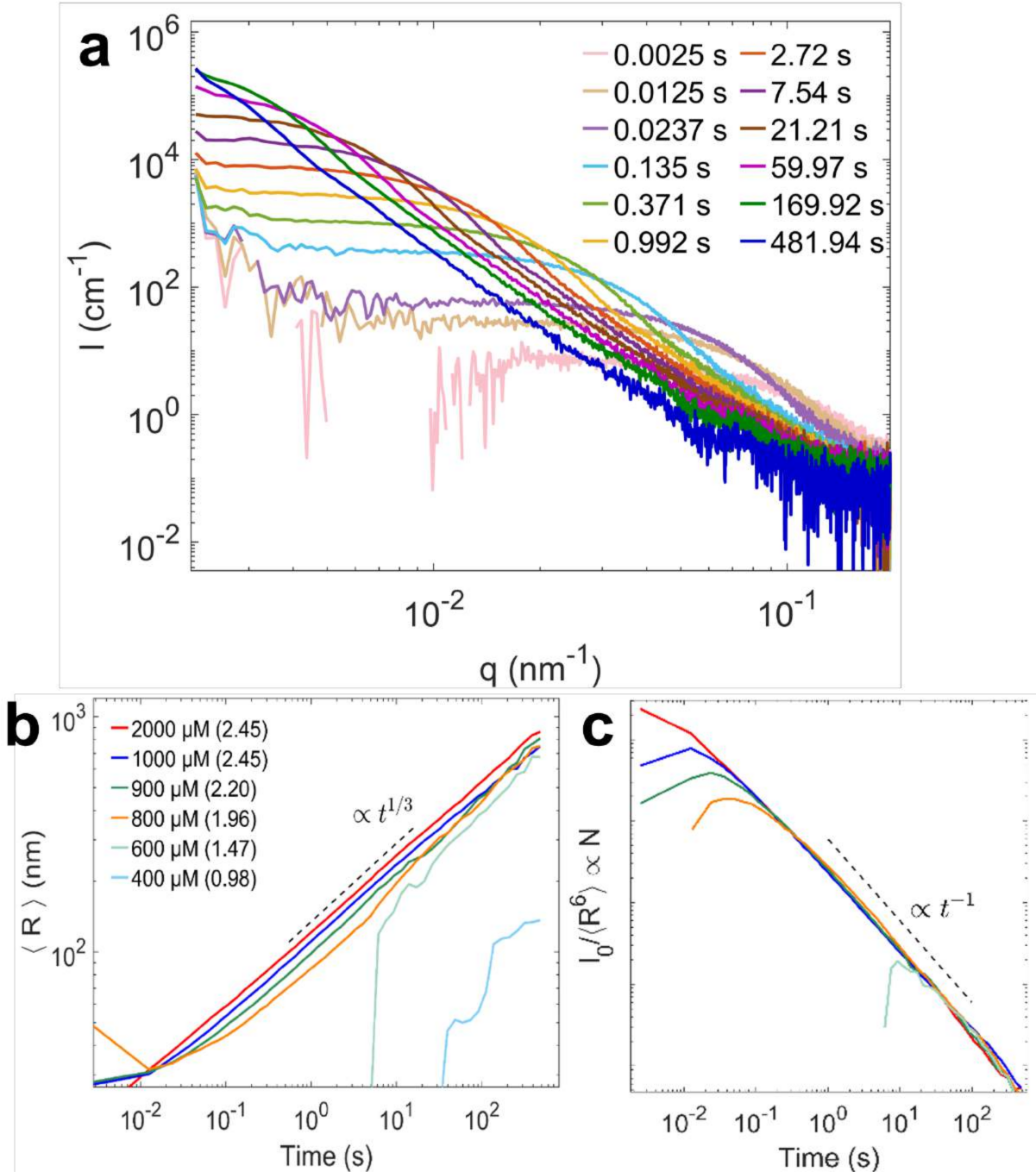}
\caption{TR-USAXS measurements under passive conditions. (a) Scattering curves at different times of a solution containing 1,000 µM of peptide and 0.5 g.L\textsuperscript{-1} of RNA. (b) Mean radius $\langle R\rangle$ of condensates as a function of time for various peptide concentrations. The numbers in brackets are the peptide-to-RNA charge ratios. $\langle R\rangle$ is estimated by fitting the scattering curves with a model of lognormally-distributed spheres. The dashed line represents a scaling law in $t^{1/3}$. (c) Arbitrarily scaled number density of condensates versus time at various peptide concentrations. The dashed line is a scaling law in $t^{-1}$. The RNA concentration in (b) and (c) is 0.5 g.L\textsuperscript{-1} except for the highest peptide concentration of 2,000 µM where the RNA concentration is 1 g.L\textsuperscript{-1}.}
\label{fig:tr-saxs_passive}
\end{figure}

Due to the Coulomb interactions underlying coacervation, growth is expected to start as early as a few milliseconds following the mixing of components, spanning lengthscales from several tens of nanometers to a few micrometers. Accordingly, TR-USAXS experiments on synchrotron source are carried out with a very large sample-to-detector distance of 31 m, and the rapid mixing is performed with a stopped-flow device that enables a dead time as short as 2.5 ms. Figure~\ref{fig:tr-saxs_passive}a depicts scattering curves measured at different times for a peptide concentration of 1,000 µM (see Fig.~\ref{fig:SI-tr-saxs_passive} for other peptide concentrations). The coacervation begins immediately after mixing and the remarkably flat Guinier plateau at low $q$-values is characteristic of the absence of interaction between condensates. Note that these measurements are repeatable from the first milliseconds up to nearly 10 min (Fig.~\ref{fig:SI-comp_growth_kinetics_1000uM}).

Since confocal microscopy clearly indicates a spherical morphology for all the condensates (Fig.~\ref{fig:phase_diagram}c), we fit each scattering curve with a model of polydisperse spheres expressed as:

\begin{equation}
I(t,q) = \frac{I_0(t)}{\langle R^6(t)\rangle} \displaystyle\int_0^{+\infty} \left[3\frac{\sin(qR)-qR\cos(qR)}{(qR)^3}\right]^2p(R;\langle R(t)\rangle,\Delta R(t))R^6dR + C_\mathrm{bkg}
\label{eqn:polydisp_spheres}
\end{equation}

\noindent where $R$ is the condensate radius and $p(R;\langle R(t)\rangle,\Delta R(t))$ the size distribution assumed here to be lognormal with mean radius $\langle R(t)\rangle$ and width $\Delta R(t)$. $C_\mathrm{bkg}$ is a constant accounting for the background. The three fitting parameters at each time step are therefore $I_0$, $\langle R\rangle$ and $\Delta R$, $\langle R^6\rangle$ being computed with the size distribution. Figure~\ref{fig:tr-saxs_passive}b shows the time evolution of $\langle R\rangle$ inferred from the scattering curves at various peptide concentrations. Strikingly, the mean radius obeys a scaling law of the form $\langle R\rangle= (Kt)^{1/3}$ -- where $K$ is a dynamic prefactor --, as early as the first few tens of milliseconds, and this behavior persists over nearly four decades in time for the highest peptide concentrations. An exponent of $1/3$ during liquid-liquid phase separation is the hallmark of either\cite{bray_theory_2002,berry_physical_2018}: (i) diffusion-limited Ostwald ripening where condensate-forming molecules migrate from small condensates to large condensates; or (ii) Brownian motion-induced coalescence during which condensates fuse to form increasingly larger condensates. In the latter case, the dynamic prefactor is expressed as $K\simeq 6k_\mathrm{B}T\phi/5\pi\eta$, with $k_\mathrm{B}$ the Botzmann constant, $T$ the temperature, $\phi$ the condensate volume fraction, and $\eta$ the solvent viscosity\cite{berry_physical_2018}. Whereas $K$ depends on the condensate volume fraction for the coalescence mechanism, it is only related to the properties of condensate-forming molecules such as the molar volume and the saturation concentration in the case of Ostwald ripening. The ratio of the measured dynamic prefactors for experiments performed with peptide concentrations of 1,000 and 2,000 µM while maintaining the peptide-to-RNA charge ratio constant, is estimated to be 1.6. Although a factor 2 is theoretically expected, the fact that $K$ increases with volume fraction suggests that coalescence mostly drives the growth kinetics. 

A quantity proportional to the number density of condensates $N$ can be deduced from the forward scattering intensity $I_0\equiv I(q\rightarrow 0)$ and the size distribution at each time since $I_0\sim N\langle V^2\rangle\sim N\langle R^6\rangle$ with $V$ the condensate volume \cite{mykhaylyk_general_2025}. Figure~\ref{fig:tr-saxs_passive}c depicts this quantity versus time and demonstrates that $N$ scales as $t^{-1}$, collapsing onto a single curve at late times irrespective of the peptide concentration. This behavior is consistent with a coalescence mechanism: the volume fraction $\phi$ remains constant over time, and applying the scaling law for coalescence, $R^3\sim\phi t\sim NR^3t$, leads to $N\sim t^{-1}$, with a prefactor that is indeed independent of $\phi$.

Additionally, Figure~\ref{fig:tr-saxs_passive}b illustrates the nucleation preceding the condensate growth. As the peptide concentration approaches the saturation concentration ($\approx$400 µM), an increasingly long induction time is observed before a sufficient signal-to-noise ratio on the scattering intensity enables to infer condensate size (see the 400 and 600 µM curves on Fig.~\ref{fig:tr-saxs_passive}b). A close examination of the SAXS curves at 400 µM reveals no detectable evolution until $t=32\text{ s}$, while within the following seconds, the scattering intensity dramatically increases. We hypothesize that prior to nucleation, each isolated RNA chain is decorated with bound peptides. This interpretation is supported by a comparison between the average scattering intensity measured prior to nucleation and the sum of the individual contributions from peptides and RNA. As the former exceeds the latter (Fig.~\ref{fig:SI-supersaturation}), peptides are indeed bound to RNA. From the forward scattering intensity, we estimate that approximately 30 peptides are associated with each RNA chain (see Supporting Information). Furthermore, fitting the average scattering intensity using a Fisher–Burford model (Fig.~\ref{fig:SI-supersaturation}) \cite{sorensen_light_2001}, which describes simple fractal aggregates, yields a mass fractal dimension of 1.4. This value indicates that the clusters adopt an extended, chain-like conformation, with a size ($R_\mathrm{g} = 13\ \text{nm}$) comparable to that of bare RNA ($R_\mathrm{g} = 11.6\ \text{nm}$).

\begin{figure}[hbt!]
\centering
\includegraphics[width=0.75\linewidth]{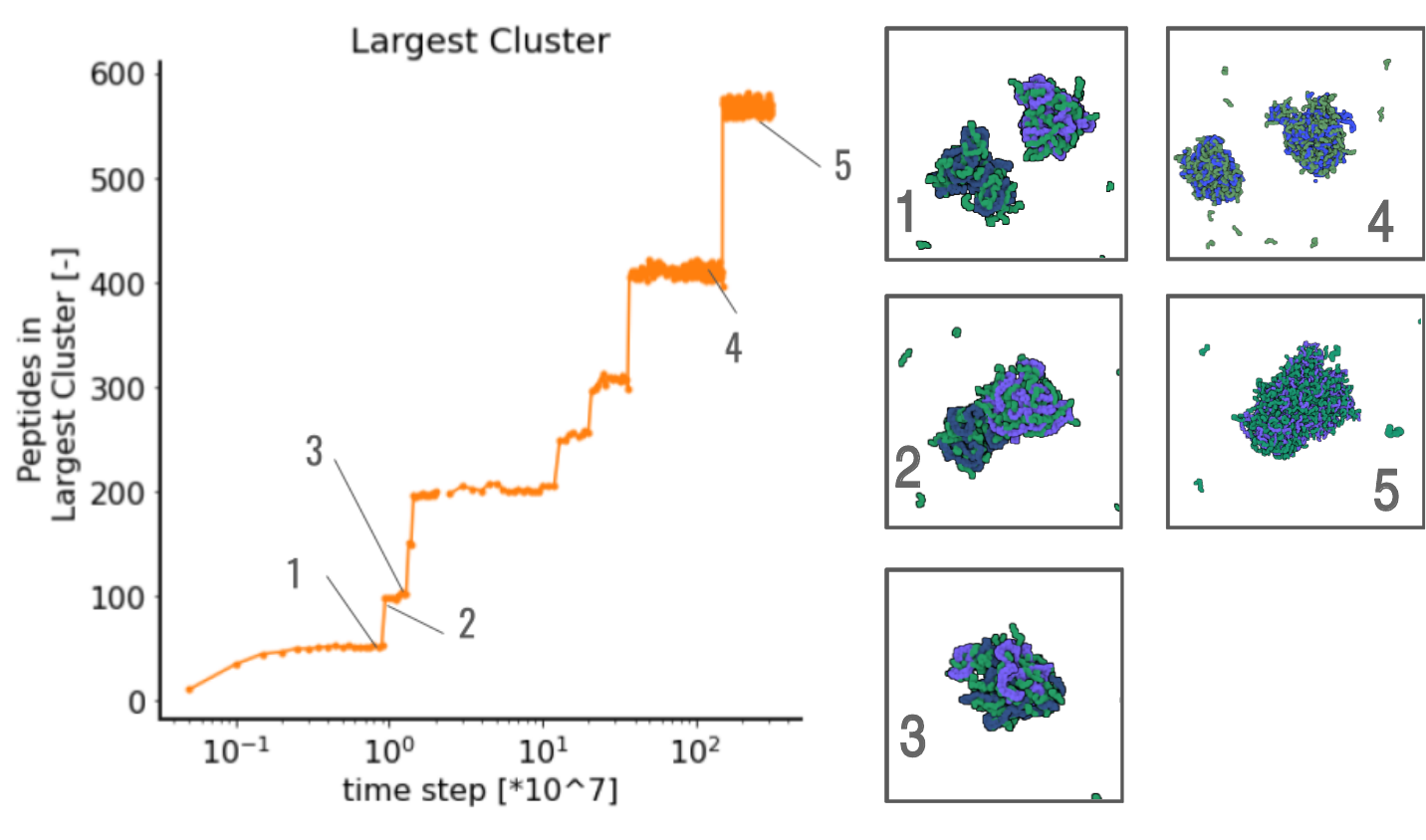}
\caption{Coarse-grained MD simulation of passively-formed condensates. The graph plots the number of peptides in the largest cluster as a function of the simulation time steps in logarithmic scale. The snapshots illustrate the coalescence events as labeled on the graph. Peptides (800 µM) are in green and RNA (0.5 g.L\textsuperscript{-1}) in purple.}
\label{fig:MD_passive}
\end{figure}

To get deeper insights into the early events at the molecular level, we perform coarse-grained molecular dynamics (MD) simulations with implicit solvent by using a recently released CALVADOS RNA force field \cite{yasuda_coarse-grained_2025}. The numerical model is able to reproduce a saturation concentration of 350 µM (Fig.~\ref{fig:SI-computed_phase_diagram}), that is, within the range of values found experimentally (Fig.~\ref{fig:phase_diagram}d; $\simeq$400 µM). Peptides and RNA are initially mixed in a simulation box, and Figure~\ref{fig:MD_passive} depicts the time evolution of the number of peptides contained in the largest peptide-RNA cluster. Very rapidly, all the peptides bind on RNA, which are thus decorated with an average of 50 peptides. Remarkably, the total number of bound peptides barely varies afterwards (Fig.\ref{fig:SI-MD_largest_cluster_all}c), confirming the role of peptide-decorated RNA as stable subunits. The peptide-RNA clusters further grow in a stepwise manner by coalescence. This is illustrated on Fig.\ref{fig:MD_passive} wherein the steps observed in the number of peptides in the largest cluster become higher over the course of the simulation. In particular, snapshots 4 and 5 show two clusters comprising 400 and 200 peptides, respectively, that is, 8 and 4 RNA chains (Fig.~\ref{fig:SI-MD_largest_cluster_all}b), respectively, merging into a single cluster. Note that fusion events are increasingly rare with the simulation time steps as the diffusion of larger clusters is much slower.

\begin{figure}[hbt!]
\centering
\includegraphics[width=0.75\linewidth]{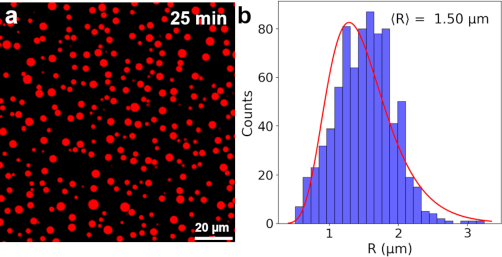}
\caption{(a) Confocal microscopy image of condensates obtained with a peptide concentration of 1,000 µM including 0.5\% of TAMRA-peptide and 0.5 g.L\textsuperscript{-1} of RNA. The image is acquired 25 min after mixing components at 37 °C. (b) Condensate radius distribution ($N=707$) inferred from a larger view of (a). The red line is a fit with a lognormal distribution.}
\label{fig:dist_passive_equilibrium}
\end{figure}

At a macroscopic level, condensates exceeding several micrometers in size cannot be probed even by USAXS. Therefore, we carry out confocal microscopy imaging by using again 0.5\% of fluorescent TAMRA-peptide. Figure~\ref{fig:dist_passive_equilibrium}a depicts a low magnification view of condensates at a peptide concentration of 1,000 µM, 25 min after mixing with RNA. The radius distribution, shown in Fig.~\ref{fig:dist_passive_equilibrium}b, resembles a lognormal distribution, as expected from a coalescence mechanism \cite{berry_physical_2018}. No macroscopic phase separation is observed, most likely as a result of Coulomb repulsion that self-regulates the growth of condensates \cite{von_hofe_multivalency_2025}.

\subsection{Network-like local structure under active condition}

\begin{figure}[hbt!]
\centering
\includegraphics[width=\linewidth]{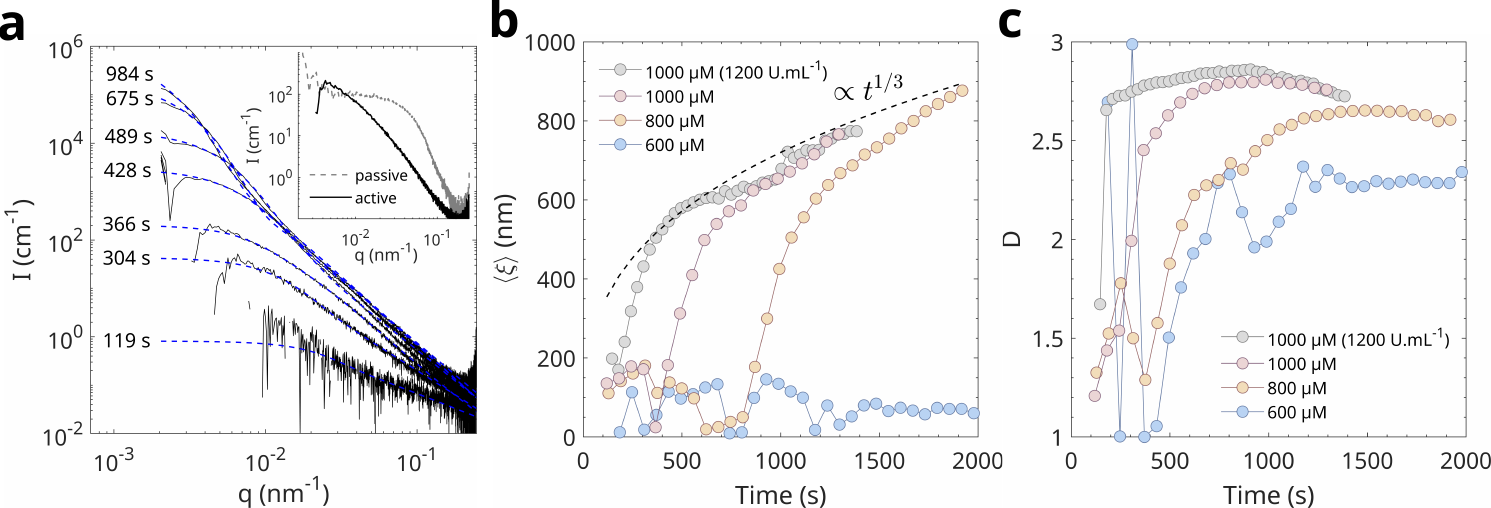}
\caption{TR-USAXS measurements under active conditions. (a) Scattering curves at various times (black solid lines) fitted with a polydisperse mass fractal model (blue dashed lines). The solution contains 1,000 µM of initially phosphorylated peptide, 0.5 g.L\textsuperscript{-1} of RNA and 800 U.mL\textsuperscript{-1} of LPP. The inset compares two scattering curves collected with peptide and RNA concentrations of 1,000 µM and 0.5 g.L\textsuperscript{-1}, respectively, under passive (gray dashed line) and active (black solid line; 800 U.mL\textsuperscript{-1} of LPP), at times where the $I_0$ values are similar. (b) Mean condensate size $\langle\xi\rangle$ as a function of time for various peptide concentrations. LPP concentration is 800 U.mL\textsuperscript{-1} except in one experiment (gray discs; 1,200 U.mL\textsuperscript{-1}). The dashed line indicates a scaling law in $t^{1/3}$. (c) Mass fractal dimension $D$ versus time for the same experimental conditions as in (b).}
\label{fig:tr-saxs_active}
\end{figure}

\noindent In the subsequent series of experiments, we examine the growth kinetics of condensates under active condition. The serine residues of the RRASLRRASL peptide are initially phosphorylated, which results in a neutral net charge. As a consequence, mixing phosphorylated peptides and RNA does not give rise to phase separation. Phase separation is instead triggered by the action of lambda phosphatase (LPP), which cleaves the phosphate groups from phosphorylated peptides. Figure~\ref{fig:tr-saxs_active}a plots scattering curves collected under the same concentrations as in Fig.~\ref{fig:tr-saxs_passive}a in the presence of 800 U.mL\textsuperscript{-1} LPP and using initially double phosphorylated peptide (see Fig.~\ref{fig:SI-tr-saxs-active} for other peptide concentrations). The condensate formation is much slower than under passive condition because phase separation is now limited by the LPP enzymatic activity. Secondly, the scattering curves in the intermediate $q$-values, i.e., between 10\textsuperscript{-2} and 10\textsuperscript{-1} nm\textsuperscript{-1}, differ markedly. The inset of Figure~\ref{fig:tr-saxs_active}a presents a comparison between two scattering curves with similar $I_0$ values. For actively-formed condensates, the decay of $I(q)$ at small and medium $q$-values is moderate ($I\sim q^{-2.4}$), which suggests a local mass fractal structure akin to that found in gels, while for passively-formed condensates, the decay slope is more consistent with a surface fractal structure. Therefore, instead of fitting the scattering curves with a model of polydisperse spheres as earlier, we apply a model of polydisperse mass fractal clusters with a compressed exponential cutoff function \cite{nicolai_static_1994,sorensen_light_2001}. The form factor of the clusters is expressed by:

\begin{equation}
F(q;\xi)=\displaystyle\int_0^{+\infty}r^{D-1}h(r;\xi)\frac{\sin (qr)}{qr}dr\big/\int_0^{+\infty}r^{D-1}h(r;\xi)dr
\end{equation}

\noindent where $D$ stands for the mass fractal dimension, and $h(r;\xi)$ is a cutoff function that describes the perimeter of the cluster of size $\xi$ and which must decay faster than any power law. The denominator ensures the normalization condition, namely, $F(q\rightarrow 0;\xi)=1$. We choose a generic exponential form for the cutoff function \cite{sorensen_light_2001}, i.e., $h(r;\xi)=e^{-(r/\xi)^\beta}$ with $\beta=3$ as it fits satisfactorily the scattering curves at late times. As before (Eq.~\ref{eqn:polydisp_spheres}), the scattering intensity is computed as a convolution product of the form factor and a lognormal size distribution defined by $p(\xi;\langle\xi(t)\rangle,\Delta\xi(t))$, so that

\begin{equation}
I(t,q)=\frac{I_0(t)}{\langle \xi^{2D(t)}(t)\rangle}\displaystyle\int_0^{+\infty}F(q;\xi)p(\xi;\langle\xi(t)\rangle,\Delta\xi(t))\xi^{2D(t)}d\xi + C_\mathrm{bkg}
\end{equation}

\noindent In practice, the scattering curves are fitted over the region $q<\text{0.1 nm\textsuperscript{-1}}$, which corresponds to lengthscales above the size of a single RNA chain. The blue dashed lines of Fig.~\ref{fig:tr-saxs_active}a represents fits to the experimental data (black solid lines), providing estimates of $I_0(t)$, $\langle\xi(t)\rangle$, $\Delta\xi(t)$ and $D(t)$ for each condition. Figure~\ref{fig:tr-saxs_active}b depicts the time evolution of $\langle\xi\rangle$ for three peptide concentrations. We observe mostly two steps: (i) Condensates with size varying around 100 nm are formed from the first minutes after mixing. These condensates display a network-like local structure as evidenced by a mass fractal dimension $D$ well below 3 (Fig.~\ref{fig:tr-saxs_active}c). After some period of maturation during which $D$ increases steadily while $\langle\xi\rangle$ remains unchanged, the condensates abruptly start to increase in size, with $\langle\xi\rangle$ seemingly converging toward the classical coarsening behavior scaling as $t^{1/3}$ (black dashed line in Fig.~\ref{fig:tr-saxs_active}b). During the latter stage, $D$ remains stable above 2.5, which indicates that the condensates are fairly compact. Notice that at 600 µM peptide concentration, no growth is observed for at least the first 2,000 s. Increasing the LPP concentration from 800 U.mL\textsuperscript{-1} to 1,200 U.mL\textsuperscript{-1} (gray discs in Figs.~\ref{fig:tr-saxs_active}b and c) shortens the maturation stage consistently with higher enzymatic activity. Finally, it is worth mentioning that although the condensate polydispersity $\Delta\xi/\langle\xi\rangle$ varies significantly during the maturation stage, it tends to a nearly constant value during the growth stage (Fig.~\ref{fig:SI-polydispersity_active}), consistently with the self-similar size distribution encountered classically in coarsening mechanisms.

\begin{figure}[hbt!]
\centering
\includegraphics[width=\linewidth]{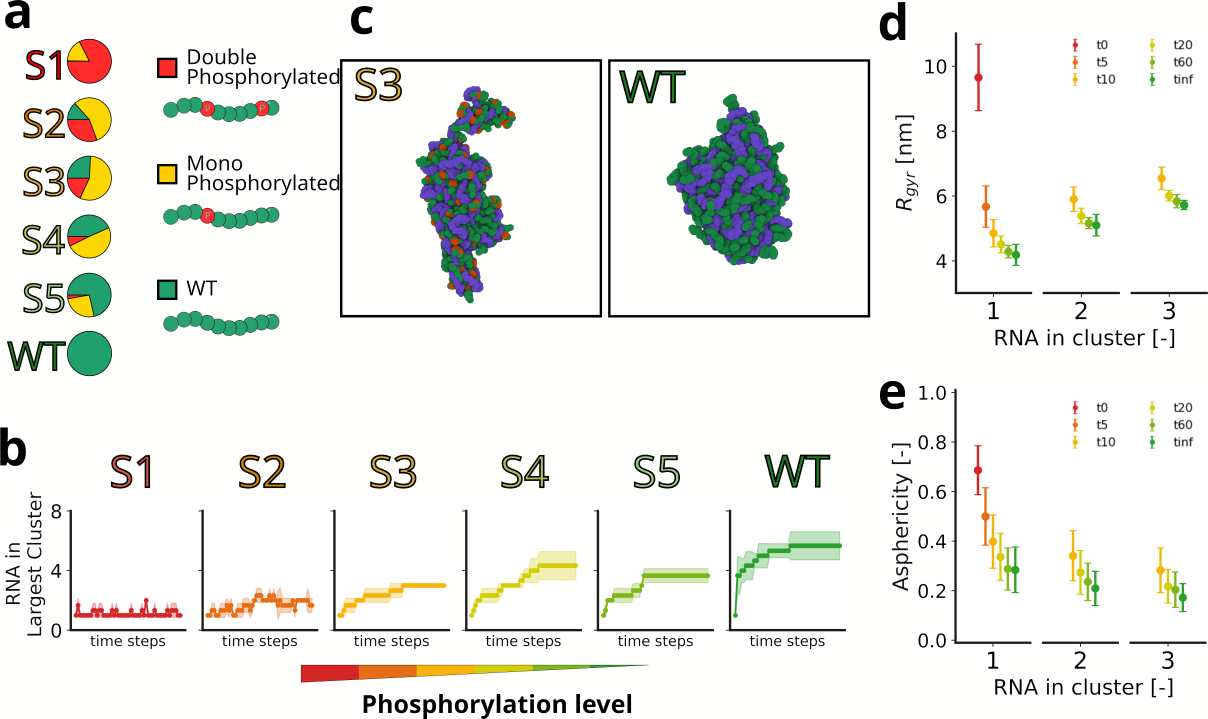}
\caption{Coarse-grained MD simulations of condensate growth mimicking active conditions. (a) Peptide compositions of the six simulated systems. (b) Time evolution of the number of RNA chains incorporated into the largest cluster for each system. Shaded areas represent standard deviations over three simulations. (c) Representative snapshots of clusters for the WT and S3 systems. (d) Radius of gyration ($R_{gyr}$) of clusters containing one, two, or three RNA chains for different peptide compositions. (e) Asphericity of clusters containing one, two, or three RNA chains for different peptide compositions.}
\label{fig:MD_active}
\end{figure}

MD simulations that explicitly account for enzymatic activity on phosphorylated peptides are not computationally feasible due to the inherently slow timescales of the reactions involved. Instead, we model cluster formation in systems containing distinct fixed peptide compositions, thereby mimicking different stages of phosphatase-driven dephosphorylation. Figure~\ref{fig:MD_active}a presents the composition of the six simulated systems, ranging from S1, which predominantly contains double phosphorylated peptides, to WT, which consists solely of non-phosphorylated peptides as set under passive conditions. The exact peptide compositions are detailed in Table~\ref{tab:SI-composition_condensate_active}. As shown in Fig.~\ref{fig:MD_active}b, decreasing the phosphorylation level enhances the ability of clusters to incorporate RNA chains within the same simulation timeframe. This behavior is expected as coacervation relies on positively charged peptides to compensate for the negative charges of RNA. More intriguingly, the final clusters display increasingly loose and heterogeneous morphologies when a substantial fraction of double or mono phosphorylated peptides is present, in contrast to the compact aggregates formed by non-phosphorylated peptides (Fig.~\ref{fig:MD_active}c). Figures~\ref{fig:MD_active}d and e quantitatively support this observation by showing that both the radius of gyration and the asphericity of the final clusters increase with peptide phosphorylation level, regardless of the number of incorporated RNA chains. In summary, higher phosphorylation levels lead to less compact and more extended cluster structures, in agreement with experimental measurements.

\begin{figure}[hbt!]
\centering
\includegraphics[width=\linewidth]{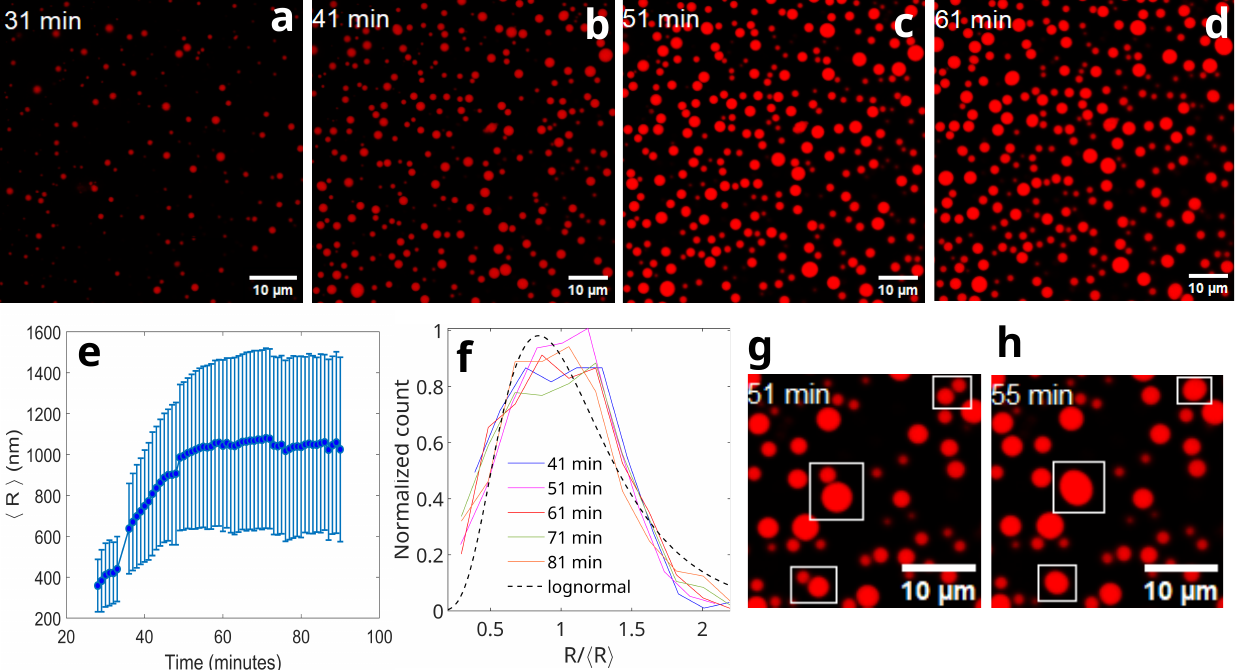}
\caption{Confocal microscopy of actively-formed condensates. Initial phosphorylated peptide concentration is 1,000 µM including 0.5\% of TAMRA-peptide, and LPP concentration is 800 U.mL\textsuperscript{-1}. (a-d) Condensate images acquired at various times after mixing components. (e) Mean condensate radius $\langle R\rangle$ versus time obtained from image analysis. Error bars are standard deviations. (f) Normalized size distribution $R/\langle R\rangle$ at various times during growth and steady-state stages. Black dashed line is a lognormal distribution plotted for comparison. (g-h) Coalescence events between condensates observed at three different locations (white boxes).}
\label{fig:confocal_active}
\end{figure}

Next, we investigate the evolution of actively-formed condensates over extended length- and timescales by confocal microscopy. Figures~\ref{fig:confocal_active}a-d depicts confocal microscopy images acquired at various times for a peptide concentration of 1,000 µM. Figure~\ref{fig:confocal_active}e represents the time evolution of the mean radius $\langle R\rangle$ and the standard deviation calculated from image analysis. $\langle R\rangle$ increases until 60 min and remains constant afterwards. Once again, the condensates attain a steady state due to self-regulation driven by Coulomb repulsion. Note that the radii deduced by confocal microscopy (Fig.~\ref{fig:confocal_active}e) are underestimated compared to those obtained by X-ray scattering (Fig.~\ref{fig:tr-saxs_active}b) because the sizes are near the resolution limit of the confocal setup. At a lower peptide concentration of 600 µM, confocal microscopy reveals that 2 h are necessary for the condensate mean radius to reach 1 µm (Fig.~\ref{fig:SI-confocal_active_600_uM}). Interestingly, the normalized size distributions, $R/\langle R\rangle$, computed at various times, are mostly superimposed, which supports the self-similarity hypothesized earlier with TR-USAXS data. Additionally, these distributions are slightly asymmetric, resembling the lognormal distribution expected for a coalescence mechanism \cite{berry_physical_2018}. This is further illustrated by Figs.~\ref{fig:confocal_active}g and h showing three coalescence events occurring over 4 min.

\subsection{Actively-formed condensates exhibit faster nanoscale dynamics}

\begin{figure}[hbt!]
\centering
\includegraphics[width=0.75\linewidth]{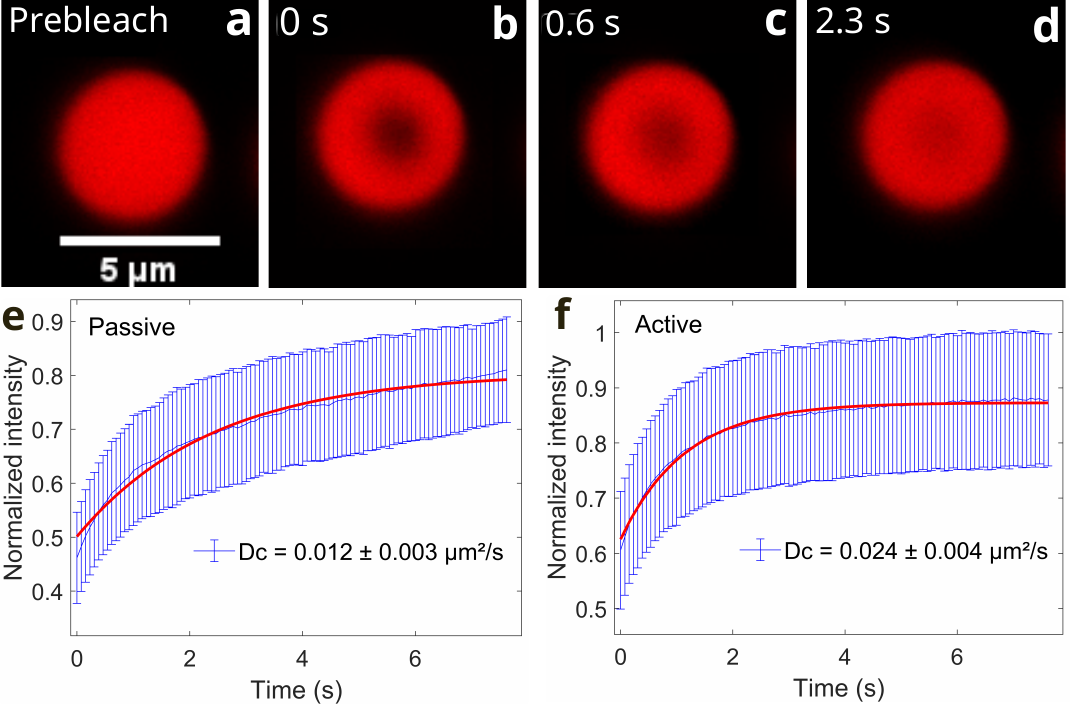}
\caption{FRAP measurements on passively- and actively-formed condensates in steady state. Peptide concentration is 1,000 µM including 0.5\% of TAMRA-peptide. For active condition, 800 U.mL\textsuperscript{-1} of LPP is used. (a-d) Confocal microscopy images of an actively-formed condensate at various times after photobleaching. Normalized fluorescence intensity versus time for passively- (e) and actively-formed (f) condensates. A dozen condensates are used for each condition. Error bars are standard deviations. The diffusivities $D_\mathrm{c}$ are deduced by fitting the data with an exponential decay function (red line).}
\label{fig:diffusion_passive_vs_active}
\end{figure}

\noindent Intrigued by the difference of local structures uncovered by X-ray scattering, we carry out fluorescence recovery after photobleaching (FRAP) measurements to estimate the diffusivity of TAMRA-peptide within both passively- and actively-formed condensates in steady state. Figures~\ref{fig:diffusion_passive_vs_active}a-d demonstrate the liquid-like nature of condensates with a fluorescence recovery obtained within just a few seconds. By integrating the fluorescence intensity over condensates and fitting the time evolution with an exponential decay function, the diffusivity of TAMRA-peptide, $D_\mathrm{c}$, is estimated (Figs.~\ref{fig:diffusion_passive_vs_active}e and f). TAMRA-peptide diffuses twice as fast within actively-formed condensates (0.024 µm\textsuperscript{2}.s\textsuperscript{-1}) as within passively-formed condensates (0.012 µm\textsuperscript{2}.s\textsuperscript{-1}). This finding agrees well with the loosely-packed, network-like local structure measured for actively-formed condensates, which confers a facilitated translational mobility to peptides.

\section{Conclusion}

\begin{figure}[hbt!]
\centering
\includegraphics[width=0.75\linewidth]{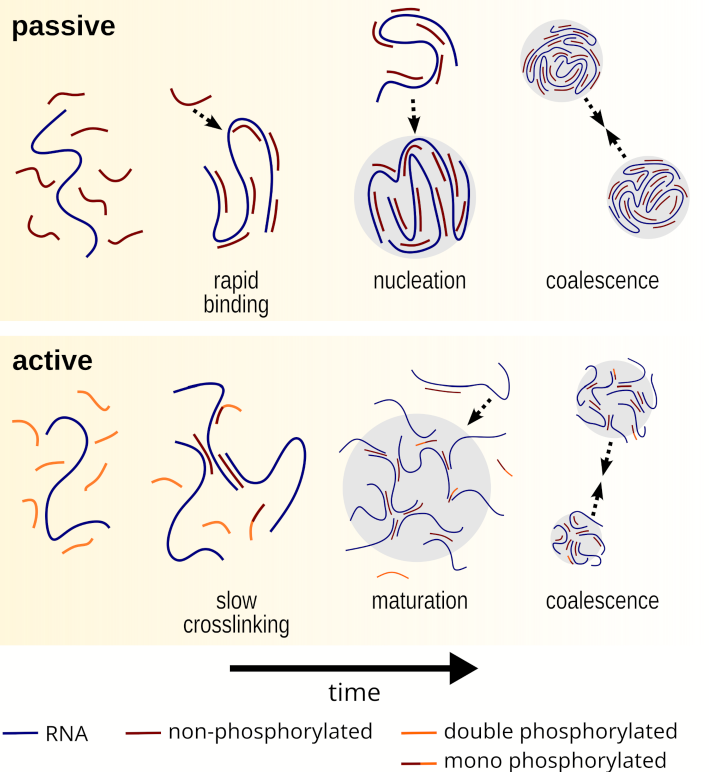}
\caption{Schematics of the growth mechanisms for passively- and actively-formed biomolecular condensates.}
\label{fig:schematics_passive_vs_active}
\end{figure}

The kinetics of the liquid–liquid phase separation leading to the formation of biomolecular condensates composed of a cationic, phosphorylatable peptide and polyU RNA -- under both passive and active conditions -- are compared using multiscale approaches that span up to four decades in time and bridge spatial scales from the molecular to the micrometer level. Based on the results, Figure~\ref{fig:schematics_passive_vs_active} proposes two growth mechanisms depending on whether enzymatic activity is involved or not. 

For passively-formed condensates, peptides bear a positive net charge and bind on RNA in less than a millisecond after mixing. Peptide-decorated RNA then serve as robust subunits and behave as an immiscible fluid in a solvent. After nucleation, spherical condensates steadily grow following the classical coarsening scaling law in $\langle R\rangle= (Kt)^{1/3}$. The prefactor $K$, as well as confocal microscopy images and MD simulations suggest a Brownian motion-induced coalescence mechanism. Eventually, coalescence is impaired by Coulomb repulsion, leading to steady state. Coalescence is preferred over ripening likely because transferring large peptide-decorated RNA from small to large condensates would be too costly in terms of free energy.

As for actively-formed condensates, double phosphorylated peptides bear a zero net charge and bind on RNA upon phosphate cleavage by LPP. Resultantly, RNA chains are slowly crosslinked by dephosphorylated peptides making up loosely-packed condensates with a network-like local structure. Recall that a double phosphorylated peptide bears two serines, each of which carries one phosphate group cleavable by LPP independently of the other. Under the action of LPP, peptides can be non-phosphorylated (two phosphate groups cleaved) or mono phosphorylated (only one phosphate group cleaved). Non-phosphorylated peptides are however required to trigger phase separation \cite{aumiller_phosphorylation-mediated_2016}. As dephosphorylation proceeds, more RNA chains are incorporated into condensates which concommitantly become more compact. After some maturation period, perhaps once non-phosphorylated peptides are supersaturated, condensate size abruptly increases upon coalescence, until steady state is reached.

Elucidating the nonequilibrium dynamics of biomolecular condensates is essential for understanding the functions and the pathological significance of membraneless organelles in the cell, as well as to design synthetic liquid-like carriers for intracellular delivery of therapeutic mRNA for instance \cite{sun_polyethylene-glycol-conjugated_2025}. Even though the phase diagram of multicomponent fluid mixtures is inherently complex, particularly in the presence of charged polymers \cite{brangwynne_polymer_2015,deviri_physical_2021,chen_size_2024}, the growth kinetics of our passively-formed condensates obey the yet-simple droplet coarsening law ($\langle R\rangle\sim t^{1/3}$) for binary mixtures. Introducing enzymatic activity delays the onset of the reaction-limited growth, but more importantly, kinetic effects modulate the local structure of condensates. A loosely-packed content may be crucial for preventing dynamical arrest that can occur with the highly charged biomolecules found in the cell nucleus \cite{galvanetto_material_2025}. In that sense, enzymatic activity may not only regulate the formation and dissolution of biomolecular condensates, but also the viscoelastic properties that govern nanoscale dynamics \cite{tanaka_viscoelastic_2022,galvanetto_material_2025}. 

A recent study has reported that kinetics govern the large-scale spatial organization of phase-separated DNA droplets \cite{wilken_spatial_2023}. In particular, fast condensation was found to promote the formation of hyperuniform structures. Although our passively-formed condensates grow on a faster timescale, we find no clear evidence of hyperuniformity, neither on scattering curves nor on confocal microscopy images (Fig.~\ref{fig:SI-structure_factor_passive_active}). Therefore, the other factors that govern the large-scale spatial organization of phase-separating fluids are yet to be identified.

At last, a number of condensates, notably transcriptional condensates, are immersed in a crowded environment that can be modeled as heterogeneous elastic medium. Recent theoretical advances have uncovered a rich diversity of condensate growth kinetics governed by local environmental stiffness \cite{meng_heterogeneous_2024}. Investigating these effects experimentally using model condensates will be an important direction for future research.

\begin{acknowledgement}

We thank Laetitia Gargowitsch for polyU characterizations, Raphaël Maire for his help on hyperuniformity and Antoine Maury for confocal microscopy observations. We have also benefited from fruitful discussions with Pau Bernado, Christine Doucet and Aurélie Fournet. This work is supported by the Agence Nationale de la Recherche (contract ANR-21-CE30-0001). G.T. acknowledges the ESRF for allocating beamtime. E.R.S is grateful to FAPESP for a visiting research scholarship at Université Paris-Saclay (\#23/018385-8).

\end{acknowledgement}

\begin{suppinfo}

Experimental methods, characterizations of RNA and peptides, supplementary figures with MD simulations, TR-USAXS data and confocal microscopy images.

\end{suppinfo}

\bibliography{sundararajan_ms_2025}

\end{document}

\section{Materials}

\begin{figure}[hbt!]
\centering
\includegraphics[width=0.5\linewidth]{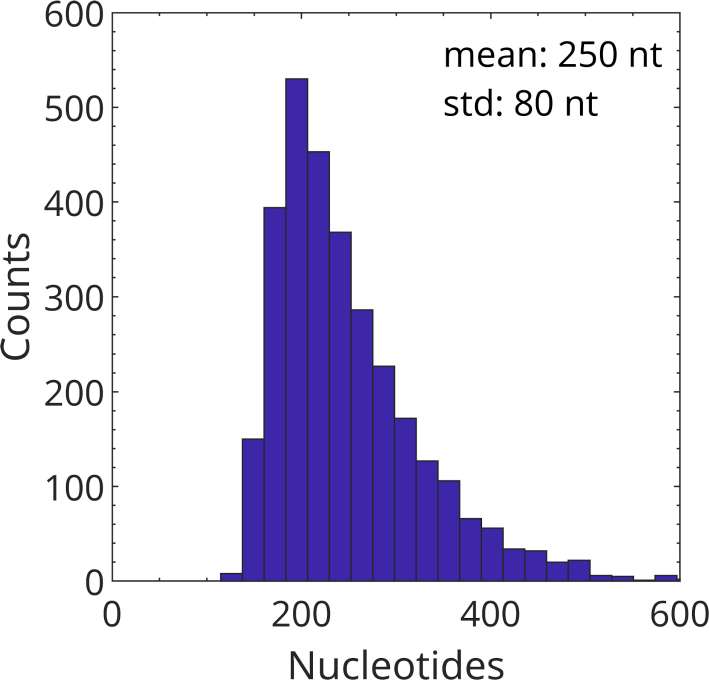}
\caption{Size distribution of polyU RNA in terms of number of nucleotides measured by mass photometry. The mean number of nucleotides is 250 and the standard deviation 80. PolyU RNA concentration is 2 nM.}
\label{fig:SI-mass_photometry_polyU}
\end{figure}

The double phosphorylated (RRApSLRRApSL), non-phosphorylated (RRASLRRASL), and N-terminus labeled 5-carboxytetramethylrhodamine (5-TAMRA) forms of the peptide were synthesized by ProteoGenix. Poly(uridylic acid) potassium salt (polyU), approximately 250 nucleotides long (Fig.~\ref{fig:SI-mass_photometry_polyU}), HEPES free acid, dithiothreitol, magnesium chloride, manganese chloride, and adenosine 5-triphosphate disodium salt hydrate were purchased from Sigma Aldrich. Lambda protein phosphatase was obtained from New England Biolabs. Poly(L-lysine)-g-poly(ethylene glycol) (PLL-g-PEG) was purchased from SuSoS (Switzerland). Buffer, RNA, and peptide stock solutions are prepared using Milli-Q water, and the buffer is subsequently filtered and autoclaved. Condensates are formed by mixing polyU RNA at a fixed concentration of 0.5 g.L\textsuperscript{-1} unless otherwise stated, with various concentrations of peptide ranging from 100-2,000 µM in a physiological buffer containing 4 mM MgCl\textsubscript{2}, 2.25 mM ATP, 1.6 mM dithiothreitol, 50 mM HEPES pH 7.4 at 37 °C. The samples are thoroughly mixed for at least 1-2 min by pipetting, which allows all components to be evenly distributed.

\section{Turbidity measurements}

Turbidity measurements are performed using an AvaSpec spectrophotometer equipped with AvaSoft software and a Quantum Northwest TC1 Peltier temperature controller at an absorbance wavelength of 500 nm at 37 °C. For the phase diagram, multiple series of experiments are conducted, varying the peptide concentration from 100 to 1,000 µM at various RNA concentrations ranging from 0.25 to 0.75 g.l\textsuperscript{-1}. The samples have a total volume of 500 µL and are incubated at 37 °C for 5 min in a quartz cuvette before measurement.  

\section{Confocal microscopy}

Confocal images are acquired using a Leica TCS SP8 laser scanning confocal inverted microscope equipped with a ×63 oil objective and an H501-T Okolab temperature-controlled stage maintained at 37 °C. The slides and coverslips are treated with O\textsubscript{2} plasma for surface activation and to ensure a sterile environment for the samples. To prevent breakage and excessive spreading of the condensates, microscopic slides and coverslips are incubated with poly(L-lysine)-g-poly(ethylene glycol) (PLL-g-PEG) at a concentration of 0.5 g.L\textsuperscript{-1} in 0.5 mL of 10 mM HEPES buffer pH 7.4 for 30 min, followed by rinsing with HEPES buffer. The PLL-g-PEG coating passivates the glass surface, preventing condensates from adhering to the slide and thereby helping maintain their spherical morphology. A small fraction ($\simeq$0.5\%) of TAMRA-conjugated peptides is incorporated into the total peptide mixture for fluorescence detection.

The slides are lined with a silicone spacer of 0.24-mm thickness. A total volume of 130 µL of sample is prepared and pipetted on microscopic slides with cavity wells, which have a depth of 0.6-0.8 mm to prevent evaporation during imaging. The samples are then covered with coverslips of 0.17-mm thickness. The excitation/emission wavelengths for TAMRA-peptide are 543 nm/569–589 nm. Time-lapse acquisition experiments are conducted, capturing images every minute over a period of 1-2 hours.

\section{Coarse-grained molecular dynamics simulations}

All molecular dynamics simulations are performed using OpenMM (v7.5) \cite{eastman_openmm_2017} in the NVT ensemble, employing a Langevin integrator with a time step of 10 fs and a friction coefficient of $0.01 \text{ ps}^{-1}$. We use the recently published CALVADOS RNA force field model \cite{yasuda_coarse-grained_2025}. In CALVADOS, RNA is modeled as two-bead-per-nucleotide model potential and includes several components: a bond term $U_b$, an angle term $U_a$, an electrostatic term $U_{\text{DH}}$, a short-range Ashbaugh-Hatch \cite{ashbaugh_natively_2008} (AH) term $U_{\text{AH}}$, and a neighbor stacking term $U_{\text{ST}}$.

The bonded potential is expressed as:

\begin{equation}
U_b = \frac{1}{2} k_b (r - r_0)^2
\end{equation}

\noindent where $k_b$ is the spring constant and $r_0$ is the equilibrium bond length. These parameters are determined separately for the backbone-backbone and backbone-base distances. For the backbone-backbone bond, the equilibrium length is $0.59 \text{ nm}$ with a spring constant of $1,400 \text{ kJ/mol/nm}^2$. For the backbone-base bond, the equilibrium length is $0.54 \text{ nm}$ with a spring constant of $2,200 \text{ kJ/mol/nm}^2$. 

The angle potential is introduced to account for the stiffness of the backbone chain:

\begin{equation}
U_a = \frac{1}{2} k_a (\theta - \theta_0)^2
\end{equation}

\noindent where $k_a$ is the spring constant, and $\theta_0$ is the equilibrium bond angle. For the backbone chain, $k_a = 4.2 \text{ kJ/mol/nm}^2$ and $\theta_0 = 3.141 \text{ rad}$. 

Short-range repulsive and attractive interactions are modeled using the Ashbaugh-Hatch (AH) potential:

\begin{equation}
U_{\text{AH}} = 
\begin{cases} 
\epsilon \left[ \left( \frac{\sigma}{r} \right)^{12} - \left( \frac{\sigma}{r} \right)^6 \right] + \epsilon (1 - \lambda) & \text{for } r \leq 2^{1/6} \sigma \\
4 \lambda \epsilon \left[ \left( \frac{\sigma}{r} \right)^{12} - \left( \frac{\sigma}{r} \right)^6 \right] & \text{for } r > 2^{1/6} \sigma
\end{cases}
\end{equation}

\noindent where $\epsilon = 0.8368 \text{ kJ/mol}$, and $\sigma$ and $\lambda$ are the effective molecular diameter and stickiness parameters, respectively. For interactions between different bead types, the combination rules are applied: $\sigma = (\sigma_i + \sigma_j)/2$ and $\lambda = (\lambda_i + \lambda_j)/2$. The AH potential is truncated and shifted at a cutoff distance of 2 nm. Neighbor stacking interactions are modeled by the AH potential with a smaller $\sigma$ to mimic strong anisotropic interactions between neighboring beads. 

The stacking potential is expressed as:

\begin{equation}
U_{\text{ST}} = n_{\text{ST}} U_{\text{AH}} (\sigma_{\text{ST}})
\end{equation}

\noindent where $n_{\text{ST}}$ is the scaling factor (typically larger than unity), and $\sigma_{\text{ST}}$ is smaller than the bead diameter to mimic strong interactions between neighbors. In this case, $\sigma_{\text{ST}} = 0.4 \text{ nm}$ and $n_{\text{ST}} = 15$. The stacking potential is also truncated and shifted at a distance of 2.0 nm, consistent with the AH potential.

Electrostatic interactions are modeled using the Debye-Hückel (DH) potential:

\begin{equation}
U_{\text{DH}} = \frac{q_i q_j}{4 \pi \epsilon_0 \epsilon_r} \frac{\exp(-r/\lambda_\mathrm{D})}{r}
\end{equation}

\noindent where $q_i$ and $q_j$ are the charges of the interacting beads (with $-e$ for backbone beads and $0$ for base beads), $\epsilon_0$ is the vacuum permittivity, $\epsilon_r$ is the dielectric constant of water set to 80, and $\lambda_\mathrm{D}$ is the Debye length.  The DH potential is truncated and shifted at a cutoff distance of 4 nm. The Debye length was selected to accurately reproduce the experimental phase diagram (ionic strength of 20 mM).

All peptides are modeled using the CALVADOS 2 force field \cite{tesei_improved_2023}. For peptide-RNA interactions, the combination rules for the AH potential are applied as described above.

\begin{figure}[hbt!]
\centering
\includegraphics[width=\linewidth]{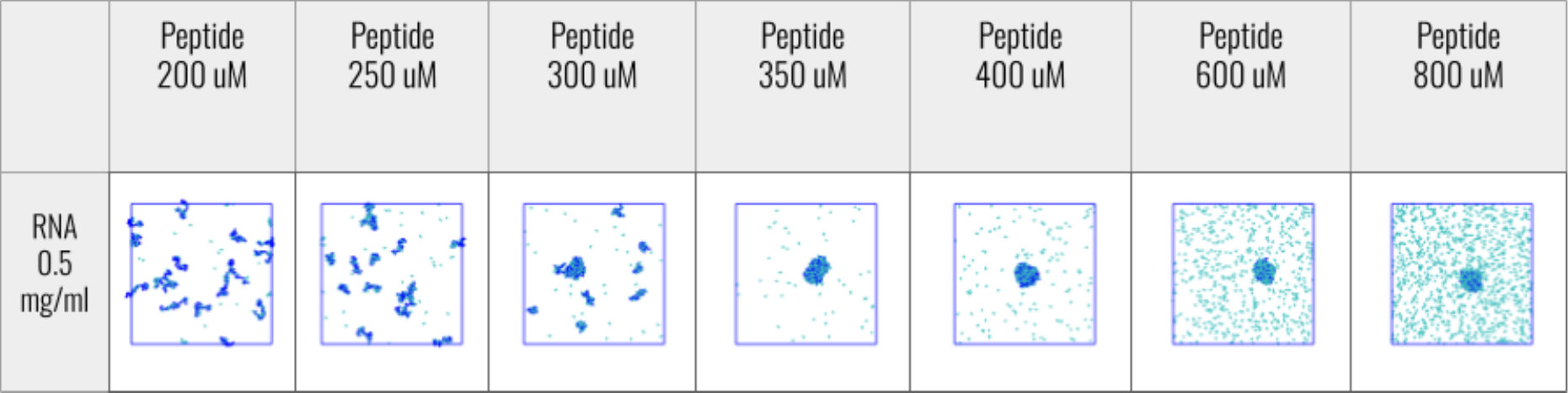}
\caption{Numerical phase diagram showing the condensation of RNA chains (purple) by peptides (green) at varying peptide concentrations. The peptide saturation concentration is around 350 µM.}
\label{fig:SI-computed_phase_diagram}
\end{figure}

\begin{figure}[hbt!]
\centering
\includegraphics[width=\linewidth]{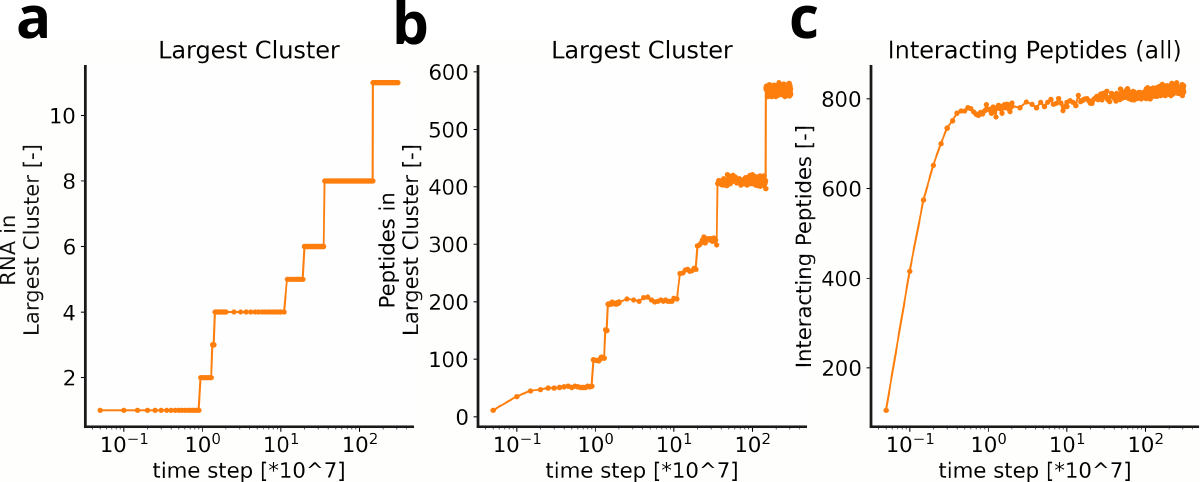}
\caption{MD simulations on the formation of peptide-RNA clusters under passive condition. (a) Time evolution of the number of RNA chains in the largest cluster. (b) Time evolution of the number of peptides in the largest cluster. (c) Total number of peptides interacting with RNA chains as a function of time. The peptide concentration is 800 µM and the RNA concentration 0.5 g.L\textsuperscript{-1}.}
\label{fig:SI-MD_largest_cluster_all}
\end{figure}

Under passive condition, simulations are performed at peptide concentrations of 200~$\mu$M, 250~$\mu$M, 300~$\mu$M, 350~$\mu$M, 400~$\mu$M, 600~$\mu$M, and 800~$\mu$M, in the presence of polyU (250 nucleotides) at a fixed concentration of 0.5~mg/mL. All simulations are carried out in a cubic simulation box with side length of 158~nm.
A phase diagram (Fig.~\ref{fig:SI-computed_phase_diagram}) as a function of peptide concentration is constructed by conducting simulations (three replicates per concentration), each initialized from a preformed peptide-RNA condensate. The saturation concentration required for condensate formation is identified to be 350~$\mu$M. To investigate the fast dynamics of cluster formation, we select a peptide concentration of 400~$\mu$M in combination with 0.5~g/L polyU. A total of 12 simulations are performed starting from a random spatial distribution of peptides and RNA molecules within the simulation box. Each simulation is run for 1~$\mu$s (corresponding to $10^8$ time steps). For each trajectory, we analyze the number of peptides and RNA chains within the largest cluster (see e.g. Fig.~\ref{fig:SI-MD_largest_cluster_all}), defined using a cutoff distance of 1~nm between residues. Three selected replicas are extended up to 30~$\mu$s to assess the system's behavior over longer timescales.

To evaluate the effect of phosphorylation on condensate behavior, simulations are performed using the latest version of the CALVADOS force field for phosphorylated disordered protein \cite{phospho2025calvados}. Three types of peptides are considered: double phosphorylated (RRApLRRApL), mono phosphorylated (RRApLRRASL), and non-phosphorylated/wild-type (WT; RRASLRRASL). For phosphorylated serine residues, charges of both -1 and -2 are tested. Simulations are carried out at a fixed peptide concentration of 800~$\mu$M. Each simulation is run for 1~$\mu$s. The relative proportions of the three peptide species are varied as detailed in Table~\ref{tab:SI-composition_condensate_active}, in order to mimic the time-dependent effect of phosphatase activity. Three replicates are performed for each system. Simulations are analyzed in terms of the number of RNA chains in the largest cluster as a function of time. Cluster geometry is evaluated by computing the gyration tensor $\mathbf{S}$ over all residues belonging to the largest cluster. Asphericity $b$ and radius of gyration $R_{gyr}$ are calculated from the eigenvalues of the gyration tensor ($\lambda_1$, $\lambda_2$, $\lambda_3$) as follows:

\begin{equation}
\begin{cases}
b = \lambda_1 - \frac{1}{2}(\lambda_2 + \lambda_3) \\
R_{gyr}^2 = \text{Tr}(\mathbf{S}) = \lambda_1 + \lambda_2 + \lambda_3
\end{cases}
\end{equation}

\begin{table}[h]
\begin{center}
\begin{tabular}{@{}lcccc@{}}
\toprule
System & WT {[}\%{]} & mono {[}\%{]} & double {[}\%{]} &  \\ \midrule
S1     & 0.3        & 17.1         & 82.6            &  \\
S2     & 13.4       & 56.0         & 30.6            &  \\
S3     & 25.9       & 56.0         & 18.1            &  \\
S4     & 43.3       & 49.3         & 7.4             &  \\
S5     & 71.5       & 25.8         & 2.8             &  \\ \bottomrule
\end{tabular}
\caption{Peptide composition for simulations mimicking different phosphorylation levels.}\label{tab:SI-composition_condensate_active}
\end{center}
\end{table}

\section{Small-angle X-ray scattering (SAXS)}

SAXS experiments are conducted at the ID02 beamline of the European Synchrotron Radiation Facility (ESRF) in Grenoble (France) at a photon energy of 12.4 keV. The 2D scattering patterns are collected using an Eiger2 4M detector (Dectris, Baden, Switzerland). The data are then radially averaged to 1D intensity profiles $I(q)$, where $q$ is the magnitude of the scattering wavevector and is expressed as $q=(4\pi/\lambda)\sin(\theta/2)$ with $\lambda$ the wavelength and $2\theta$ the scattering angle.

The sample-to-detector distance is set to 31 m to explore $q$-values ranging from $2\times 10^{-3}$ to $0.25\text{ nm}^{-1}$, enabling us to gain more insights into the large-scale structures of biomolecular condensates. For measuring the individual components, the sample-to-detector distance is set to 3 m for RNA and 1 m for peptide, reaching a $q$ range of $2\times 10^{-2}$ to $2\text{ nm}^{-1}$. All the measurements are carried out at 37 °C.

The sample is placed in a flow-through quartz capillary cell with a diameter of 2 mm. For each measurement, around 100 µL of sample solution is thoroughly mixed by repeated pipetting for about one minute and is injected into the capillary. The capillary is thoroughly cleaned using Hellmanex, ethanol, deionized water, buffer, and dried using compressed air between different samples. To ensure the samples are washed away from the capillary, the background sample (buffer) is often measured between samples to make sure it remains consistent with previously measured buffers.

For fast kinetics, time-resolved ultra-small-angle X-ray scattering (TR-USAXS) experiments are performed using an SFM-4000 stopped-flow device (Bio-Logic Science Instruments, France) to capture the early timescales of condensate formation. Stock solutions of RNA, peptide, and buffer are loaded into syringes, and the required final solutions are injected into the stopped-flow capillary cell with a diameter of 1.2 mm at a total flow rate of 4.2 or 6.25 mL.s\textsuperscript{-1}, resulting in a dead time of 2.5 ms. For each kinetic measurement, 50 successive frames are captured with an exposure time of 5 ms per frame. The intervals between frames followed a geometric progression with a ratio of 1.23, and each kinetic measurement is repeated at least twice to ensure reliability.

For both SAXS and TR-USAXS, the scattering intensities are scaled to absolute intensities. Further data processing, such as buffer subtraction and averaging, is performed using SAXSUtilities software \cite{saxsutilities2_2021}. TR-USAXS curves are then fitted with a polydispersity sphere model with lognormal size distribution to extract the radius of the condensates using SASView package \cite{sasview_2023}. The fits with a polydisperse mass fractal model are carried out with custom-made routines in MATLAB\textsuperscript{\textregistered}.

\begin{figure}[hbt!]
\centering
\includegraphics[width=0.5\linewidth]{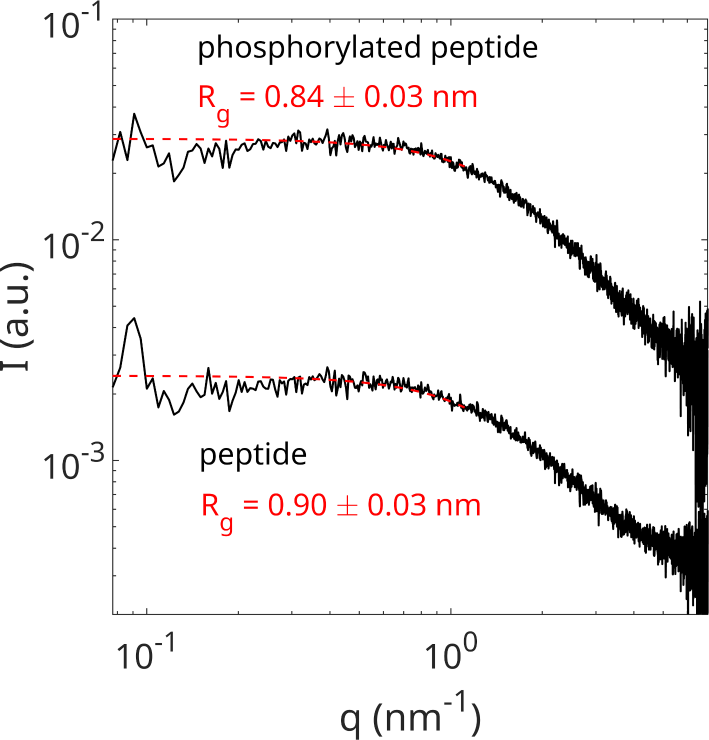}
\caption{SAXS data (black lines) of RRASLRRASL peptide and its phosphorylated form at 2.95 mM and 3.4 mM, respectively. The red dashed lines are Guinier fits plotted for $qR_\mathrm{g}<1$. The curves are shifted for clarity.}
\label{fig:SI-saxs_peptide}
\end{figure}

\begin{figure}[hbt!]
\centering
\includegraphics[width=0.5\linewidth]{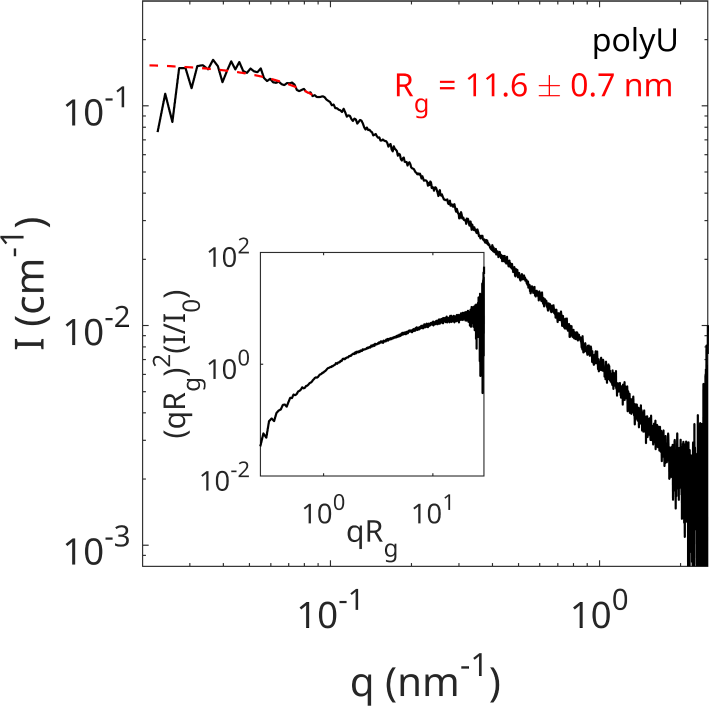}
\caption{SAXS data (black line) of polyU RNA at 1.5 g.L\textsuperscript{-1}. The red dashed line is a Guinier fit plotted for $qR_\mathrm{g}<1$. The inset is a Kratky representation of the same data.}
\label{fig:SI-saxs_polyU}
\end{figure}

\begin{figure}[hbt!]
    \centering
    \includegraphics[width=1\linewidth]{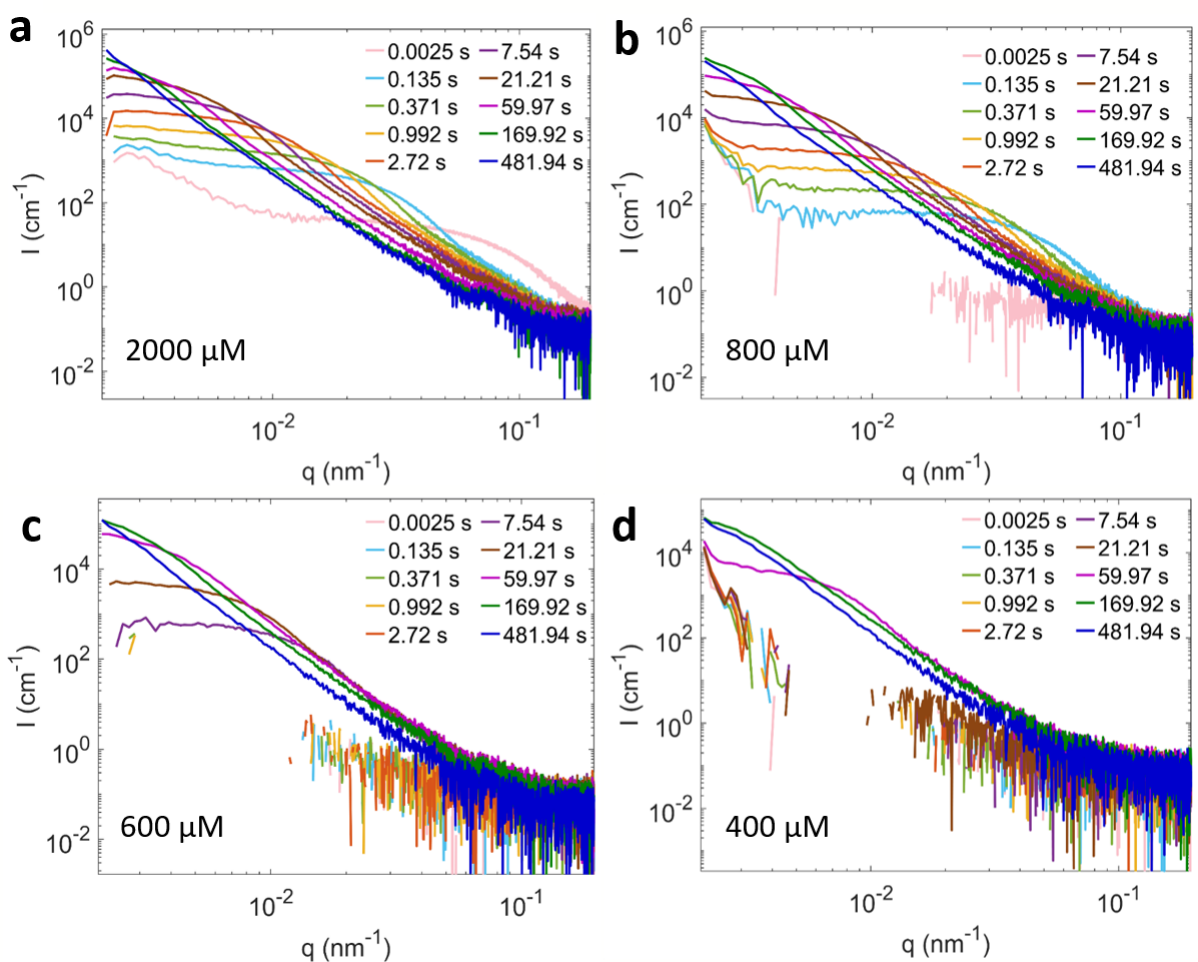}
    \caption{TR-USAXS curves showing the growth kinetics of passively-formed condensates at peptide concentrations ranging from 2,000 to 400 µM (a to d). Each curve represents a distinct time from 2.5 ms to 481.94 s. RNA concentration is maintained at 0.5 g.L\textsuperscript{-1} for all peptide concentrations, except at 2,000 µM peptide, where it is increased to 1 g.L\textsuperscript{-1}. The progressive leftward shift of the Guinier region at low $q$-values indicates condensate growth.}
    \label{fig:SI-tr-saxs_passive}
\end{figure}

\begin{figure}[hbt!]
\centering
\includegraphics[width=0.5\linewidth]{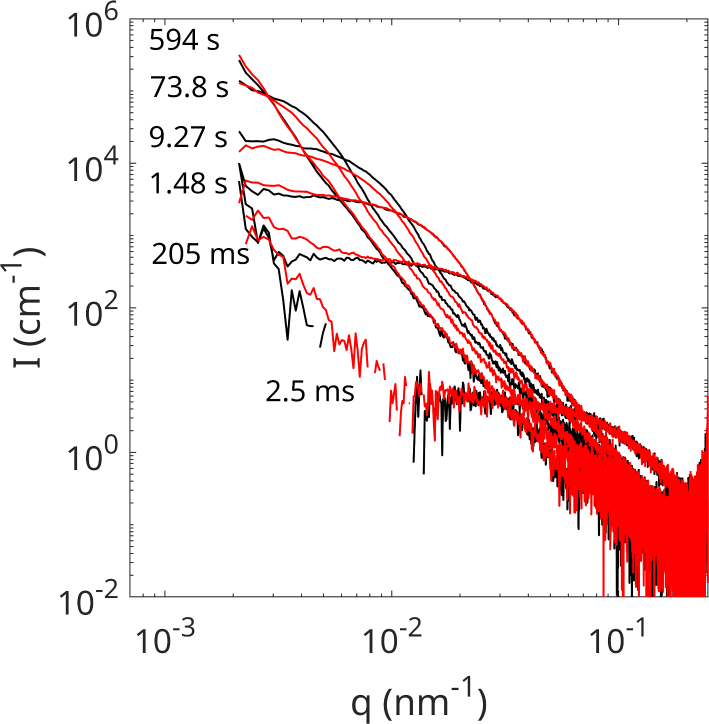}
\caption{Comparison between two consecutive sets of TR-USAXS data (in black and red) at selected times, under passive condition. Both experiments are performed by mixing 1,000 µM of peptide with 0.5 g.L\textsuperscript{-1} of RNA.}
\label{fig:SI-comp_growth_kinetics_1000uM}
\end{figure}

\begin{figure}[hbt!]
\centering
\includegraphics[width=0.5\linewidth]{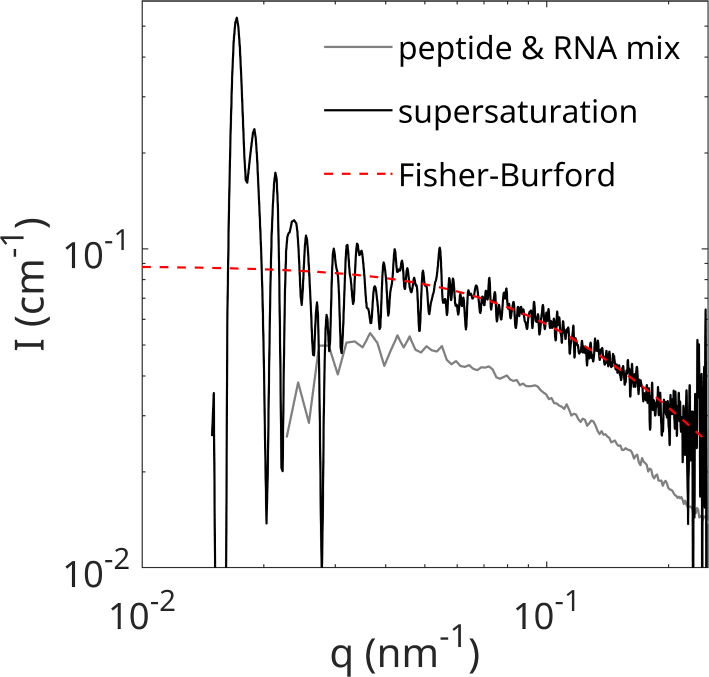}
\caption{Supersaturated peptide-decorated RNA with 400 µM of peptide and 0.5 g.L\textsuperscript{-1} of RNA. (Black solid line) Average scattering curves measured during induction. (Gray solid line) Scattering curve calculated as the sum of peptide and RNA contributions. (Red dashed line) Fisher-Burford fit with a mass fractal dimension of 1.4 and a radius of gyration of 13 nm.}
\label{fig:SI-supersaturation}
\end{figure}

At the supersaturation, before condensate formation, RNA chains are not aggregated and each of them is assumed to be decorated with a number $n$ of peptides. The forward scattering intensity $I_0$ can be written as:

\begin{equation}
I_0=c_\mathrm{RNA}\left( n\Delta b_\mathrm{p}+\Delta b_\mathrm{RNA}\right)^2+(c_\mathrm{p}-nc_\mathrm{RNA})\Delta b_\mathrm{p}^2
\label{eq:supersaturation}
\end{equation}

\noindent with $c_\mathrm{RNA}$ and $c_\mathrm{p}$ are the molar RNA and peptide concentrations, respectively, while $\Delta b_\mathrm{RNA}^2$ and $\Delta b_\mathrm{p}^2$ are the excess scattering lengths related to RNA and peptide, respectively. The first term of Eq.~\ref{eq:supersaturation} arises from the RNA chains decorated with $n$ peptides, whereas the second term accounts for the remaining free peptides. From separate measurements (Figs.~\ref{fig:SI-saxs_polyU} and \ref{fig:SI-saxs_peptide}), we find $\Delta b_\mathrm{RNA}=0.0892\text{ SI}$ and $\Delta b_\mathrm{p}=9.057\times10^{-4}\text{ SI}$ for scattering intensities in cm\textsuperscript{-1} and molar concentrations in µM. By solving Eq.~\ref{eq:supersaturation}, the number of bound peptides per RNA chain reads\cite{chevreuil_nonequilibrium_2018}

\begin{equation}
n=\displaystyle\sqrt{\Gamma^2+\frac{I_0-I_0^*}{c_\mathrm{RNA}\Delta b_\mathrm{p}^2}}-\Gamma
\end{equation}

\noindent where $\Gamma=\Delta b_\mathrm{RNA}/\Delta b_\mathrm{p}-1/2$ and $I_0^*=c_\mathrm{RNA}\Delta b_\mathrm{RNA}^2+c_\mathrm{p}\Delta b_\mathrm{p}$. With $I_0$ estimated from the scattering curve during induction (Fig.~\ref{fig:SI-supersaturation}), we obtain $n=30\pm 3$. The scattering curve is further fitted with a Fisher-Burford model\cite{sorensen_light_2001} that provides a simple approximation for mass fractal clusters, i.e.,

\begin{equation}
I(q)=I_0\left(1+\frac{2}{3D}q^2R_\mathrm{g}^2\right)^{-D/2}
\end{equation}

\noindent where $D$ stands for the mass fractal dimension. We obtain $D=1.4$ and $R_\mathrm{g}=13\text{ nm}$, which indicates that peptide-decorated RNA at the supersaturation exhibit an extended chain-like conformation with an overall size close to that of bare RNA.

\begin{figure}[hbt!]
    \centering
    \includegraphics[width=1\linewidth]{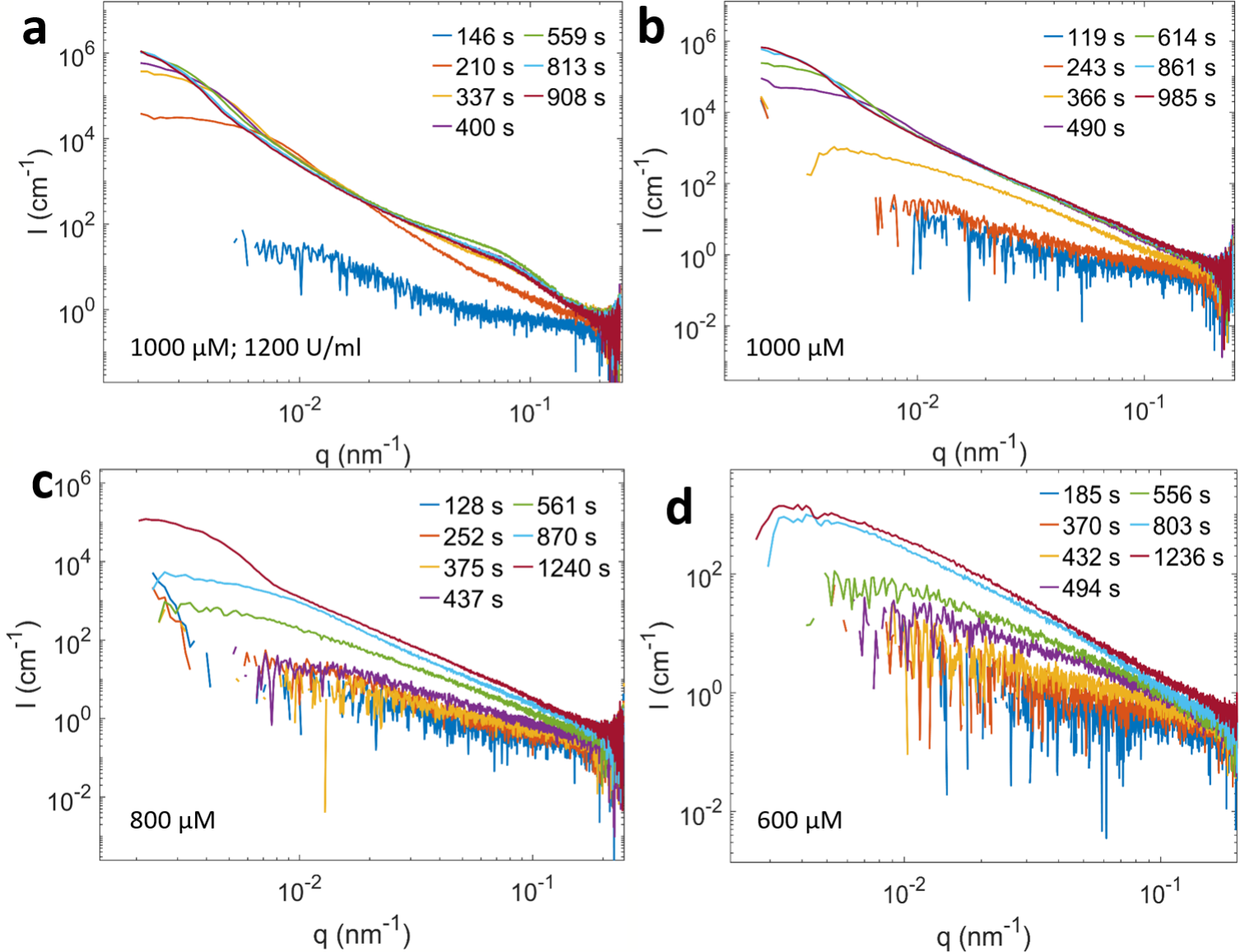}
    \caption{TR-USAXS curves showing the growth kinetics of actively-formed condensates with double phoshorylated peptide concentrations ranging from 1,000 to 600 µM (a to d). RNA concentration is maintained at 0.5 g.L\textsuperscript{-1} for all samples and LPP concentration is set to 800 U.mL\textsuperscript{-1} for all samples (b-d), except in (a), where it is increased to 1,200 U.mL\textsuperscript{-1}. The bump at $q\simeq\text{0.8 nm\textsuperscript{-1}}$ on panel (a) is due to an instrumental artifact that could not be corrected \emph{a posteriori}.}
    \label{fig:SI-tr-saxs-active}
\end{figure}

\begin{figure}[hbt!]
\centering
\includegraphics[width=0.5\linewidth]{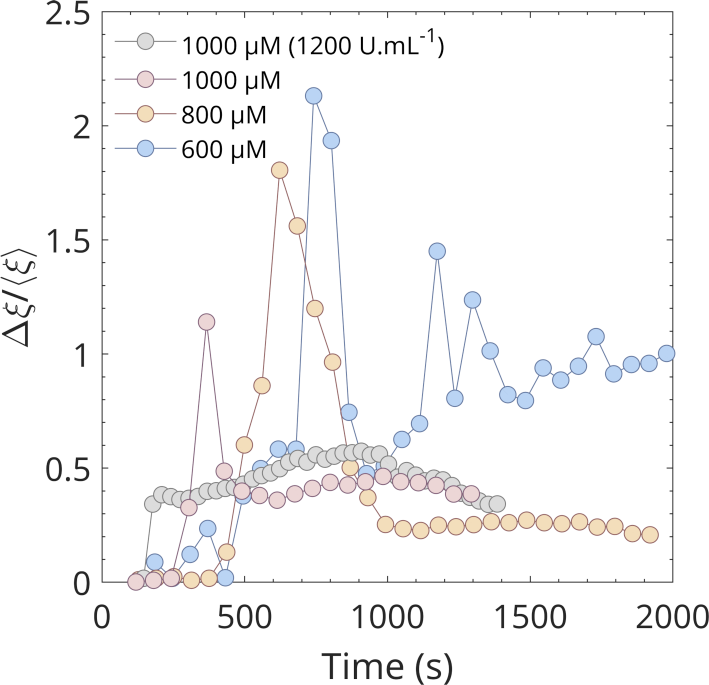}
\caption{Time evolution of the polydispersity $\Delta\xi/\langle\xi\rangle$ under active conditions. LPP concentration is 800 U.mL\textsuperscript{-1} except in one experiment (gray discs) where it is 1,200 U.mL\textsuperscript{-1}.}
\label{fig:SI-polydispersity_active}
\end{figure}

\begin{figure}[hbt!]
    \centering
    \includegraphics[width=1\linewidth]{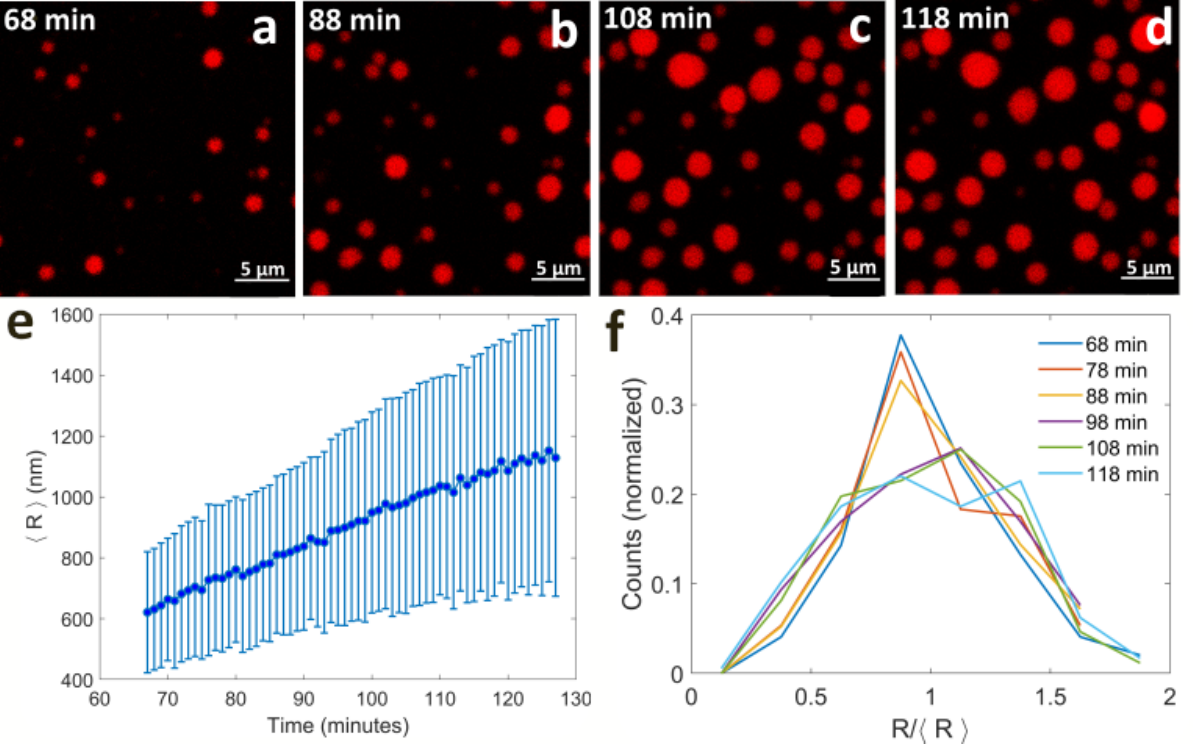}
    \caption{(a–d) Confocal microscopy images of condensates formed under active condition, taken at various times after mixing. The initial samples contain 600 µM of double phosphorylated peptide, $\sim 0.5\%$ TAMRA-peptide, 0.5 g.l\textsuperscript{-1} RNA and 800 U.mL\textsuperscript{-1} LPP. (e) Mean condensate radius $\langle R\rangle$ as a function of time, obtained from image analysis. Error bars represent standard deviations. (f) Normalized size distribution normalized at different stages of condensate growth.}
    \label{fig:SI-confocal_active_600_uM}
\end{figure}

\section{Fluorescence Recovery After Photobleaching (FRAP)}

Fluorescence recovery after photobleaching (FRAP) experiments are conducted using a Leica TCS SP8 laser scanning confocal inverted microscope equipped with a $\times$63 oil objective and an H501-T Okolab temperature-controlled stage maintained at 37 °C. Individual condensates are photobleached for 2 s. A total of 100 frames are captured with a time interval of 77 ms to monitor fluorescence recovery. Approximately 15 condensates are analyzed under passive and active conditions.

The confocal images are processed using a custom ImageJ macro that normalizes fluorescence intensity within the bleached region against a control region. The macro extracts mean intensities from user-defined regions of interest (ROIs). The extracted intensity values are plotted over time and fitted with an exponential decay function. The characteristic time $\tau$ is extracted from the fit, and the diffusion coefficient $D_\mathrm{c}$ is calculated using the formula:

\begin{equation}
D_\mathrm{c} = \frac{w^2}{4\tau}
\end{equation}

\noindent where $w$ is the radius of the bleached area.

\section{Large-scale structure factor}

\begin{figure}[hbt!]
\centering
\includegraphics[width=\linewidth]{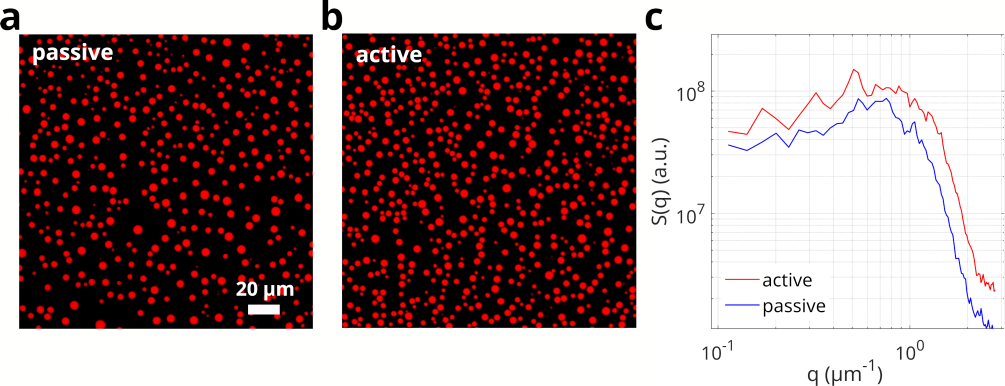}
\caption{Confocal microscopy images of condensates with peptide concentration of 1,000 µM under passive (a) and active (b) conditions. Both images are taken in steady state, when the size distribution no longer evolves significantly. (c) Structure factor $S(q)$ calculated on the two preceding images under passive (blue) and active (red) conditions.}
\label{fig:SI-structure_factor_passive_active}
\end{figure}

The analysis begins by converting each confocal microscopy image (Figs.~\ref{fig:SI-structure_factor_passive_active}a and b) to a grayscale format to simplify intensity processing. To reduce boundary artifacts and minimize the influence of interfacial regions, a radial mask is applied to the image \cite{wilken_spatial_2023}. This mask accounts for the lack of periodicity in the image and ensures that only the central region of the sample image contributes to the analysis. Next, a two-dimensional Fourier transform is performed on the masked image. This transformation converts the spatial intensity distribution into reciprocal space, yielding the intensity structure factor, which reflects the spatial correlations within the sample at a given $q$-value. Due to the isotropy of the system, a radial average of the two-dimensional structure factor is computed to obtain a one-dimensional representation of the scattering intensity. This averaging process collapses the angular information, resulting in a structure factor profile, $S(q)$. The characteristic hyperuniform scaling $S(q) \sim q^2$ at small $q$-values, observed in phase-separated DNA droplets \cite{wilken_spatial_2023} and indicative of suppressed long-wavelength density fluctuations, is not evident in this case. Instead, $S(q)$ appears to plateau at low $q$ (Fig.~\ref{fig:SI-structure_factor_passive_active}c), showing no signs of hyperuniformity in the system.

\clearpage

\bibliography{sundararajan_ms_2025}